\documentclass[aps,showpacs]{revtex4}
\usepackage{graphicx}

\newcommand{\be}{\begin{equation}}
\newcommand{\ee}{\end{equation}}
\newcommand{\ba}{\begin{eqnarray}}
\newcommand{\ea}{\end{eqnarray}}
\newcommand{\ban}{\begin{eqnarray*}}
\newcommand{\ean}{\end{eqnarray*}}

\begin{document}

\title{Coulomb Effects in Femtoscopy}

\author{Rados\l aw Maj\footnote{radmaj@ujk.kielce.pl}}

\affiliation{Institute of Physics, Jan Kochanowski University, 
ul.~\'Swi\c etokrzyska 15, PL-25-406 Kielce, Poland}

\author{Stanis\l aw Mr\' owczy\' nski\footnote{mrow@fuw.edu.pl}}

\affiliation{Institute of Physics, Jan Kochanowski University, 
ul.~\'Swi\c etokrzyska 15, PL-25-406 Kielce, Poland \\
and So\l tan Institute for Nuclear Studies, 
ul. Ho\.za 69, PL-00-681 Warsaw, Poland}

\date{September 28, 2009}

\begin{abstract}

The correlation function of two identical particles - pions or kaons -
interacting via Coulomb potential is computed. The particles are
emitted from an anisotropic particle's source of finite lifetime. 
In the case of pions, the effect of halo is taken into account as 
an additional particle's source of large spatial extension. The 
relativistic effects are discussed in detail. The Bowler-Sinyukov 
procedure to remove the Coulomb interaction is carefully tested. In
the absence of halo the procedure is shown to work very well even for 
an extremely anisotropic source. When the halo is taken into account 
the free correlation function, which is extracted by means of the 
Bowler-Sinyukov procedure, is distorted at small relative momenta 
but the source parameters are still correctly reproduced. 

\end{abstract}

\pacs{25.75.-q, 25.75.Gz}

\maketitle


\section{Introduction}


The correlation functions of two particles with small relative
momenta provide information about space-time characteristics of
particle's sources in high-energy nucleus-nucleus collisions, see the
review articles
\cite{Wiedemann:1999qn,Heinz:1999rw,Lisa:2005dd,Chajecki:2009zg}.
Within the standard femtoscopy, one obtains parameters of 
a particle's source, comparing the experimental correlation functions 
to the theoretical ones which are calculated in a given model. Such 
an analysis can be performed for pairs of non-identical or identical 
particles. In the former case, the correlation appears due to 
inter-particle interaction while in the latter one the interaction 
is combined with the effects of quantum statistics. Since we usually 
deal with electrically charged particles, observed two-particle 
correlations are strongly influenced by the Coulomb interaction. The 
effect of the Coulomb force is usually eliminated from experimental 
data by means of the so-called Bowler-Sinyukov procedure 
\cite{Bowler:1991vx,Sinyukov:1998fc}. And then, the correlation 
function, which is obtained in such a way from experimental data, is 
compared to the theoretical correlation function of two non-interacting 
particles. The comparison provides parameters of the source of particles.

The femtoscopy was applied to a large volume of experimental data 
on nucleus-nucleus collisions at SPS energy as summarized in 
\cite{Alt:2007uj}. The spatial size of particle's 
sources appeared to be comparable to the expected size a fireball 
created in nucleus-nucleus collisions while the emission time of 
particles was significantly shorter. It was predicted that at RHIC 
energies the emission time would be significantly longer due to the 
long lasting hydrodynamic evolution of the system created at the 
early stage of nucleus-nucleus collisions 
\cite{Rischke:1995cm,Rischke:1996em}. To a big surprise the 
experimental data obtained at RHIC 
\cite{Adler:2001zd,Adcox:2002uc,Adler:2004rq,Adams:2004yc} show 
a very little change of the space-time characteristics of 
a fireball when compared to the SPS data. In particular, the emission 
time of particles appeared to be as short as 1 fm/c. Because of this 
surprising result, which is now known as the `HBT  Puzzle'
\cite{Gyulassy:2001zv,Pratt:2003ij}, a reliability of the femtoscopy 
was questioned.

As an alternative to the standard femtoscopy, the method of imaging 
\cite{Brown:1997ku,Brown:2005ze} was developed. Within this method 
one obtains the source function not referring to its specific 
parametrization but directly inverting the correlation function. 
The procedure of inversion takes into account the effect of quantum 
statistics as well as that of inter-particle interaction. The 
one-dimensional and three-dimensional imaging was successfully 
applied to experimental data, see 
\cite{Panitkin:2001qb,Chung:2002vk} and 
\cite{Adler:2006as,Afanasiev:2007kk,Alt2008}, respectively.
The method provides essentially model independent information on 
the source space-time sizes but modeling is still needed to deduce 
the emission time which is coupled to spatial parameters of the 
source. Therefore, the imaging has not much helped to resolve the 
`HBT puzzle'.

Very recently it has been shown that hydrodynamic calculations can be 
modified to give quite short emission times of produced particles 
\cite{Broniowski:2008vp}, see also \cite{Pratt:2008bc,Sinyukov:2009xe}. 
Specifically, the initial condition needs to be changed to speed up 
formation of the transverse collective flow and the first order phase 
transition from quark-gluon plasma to hadron gas should be replaced 
by the smooth cross-over. Another solution of the `HBT Puzzle' assumes 
an incomplete equilibrium of quark-gluon plasma \cite{Gombeaud:2009fk}. 
Although the `HBT Puzzle' is resolvable now, if not resolved in
\cite{Broniowski:2008vp,Pratt:2008bc,Sinyukov:2009xe,Gombeaud:2009fk},
it is still of interest to quantitatively check the femtoscopy method, 
to be sure that experimentally obtained source parameters are indeed 
reliable. This is the aim of our study which is mainly focused on the 
Coulomb effects. Our preliminary results were presented  in 
\cite{Maj:2007qw} but, unfortunately, some errors appeared in this 
publication.

The Bowler-Sinyukov correction procedure, which is used to eliminate 
the Coulomb interaction from the experimental data, assumes that the 
Coulomb effects can be factorized out. The correction's factor is 
calculated for a particle's source which is spherically symmetric and 
has zero lifetime. We examine the procedure applying it to the computed 
Coulomb correlation functions of identical particles coming from 
anisotropic sources of finite lifetime. Azimuthally asymmetric sources,
which appear in azimuthally sensitive femtoscopy 
\cite{Adams:2003ra,Lisa:2003ze}, are also studied. We treat the computed 
Coulomb correlation functions as experimentalists treat the 
measured correlation  functions. Thus, we extract the correlation 
function which is supposed to be free of the Coulomb interaction. 
However, in contrast to the situation of experimentalists we know 
actual parameters of particle sources which can be compared to the 
extracted ones. Our analysis is somewhat similar to that presented 
in \cite{Kisiel:2006is} where the Coulomb correlation functions 
were computed by means of Monte Carlo event generator and it was
claimed that the procedure of removal of the Coulomb effects works
well. Our analysis is more detailed and it is based on mostly
analytical calculations. 

The correlation function of two identical non-interacting bosons is
expected to be equal to 2 for vanishing relative momentum of the two
particles. The correlation functions extracted from experimental data
by means of the procedure, which is supposed to remove the Coulomb
interaction, do not posses this property. The correlation function
at zero relative momentum is significantly smaller than 2. This fact 
is usually explained referring to the concept of halo \cite{Nickerson:1997js}. 
It assumes that only a fraction of observed particles comes from the 
fireball while the rest originates from the long living resonances. 
Then, we have two sources of particles: the fireball and the halo 
with the radius given by the distance traveled by long living 
resonances. The complete correlation function, which includes 
particles from the fireball and the halo, equals 2 at exactly 
vanishing relative momentum. However, the correlation of two particles 
coming from the halo occurs at a relative momentum which is as small 
as the inverse radius of the halo. Since experimental momentum 
resolution is usually much poorer and such small relative 
momenta are not accessible, the correlation function is claimed to be 
less than 2 for {\em effectively} vanishing relative momentum. We 
carefully study the effect of halo and, in particular, we test how 
the Bowler-Sinyukov correction procedure works in the presence of halo.

We discuss in detail how to compute the Coulomb correlation functions. 
We pay particular attention to relativistic effects which, in our 
opinion, are not clearly exposed in literature. We start with 
the nonrelativistic Koonin formula \cite{Koonin:1977fh} 
because of its rather transparent physical meaning. A more formal 
derivation of the correlation function, which follows the studies 
\cite{Lednicky:1981su,Lednicky:1996hp,Lednicky:2005tb},
is sketched in the Appendix A. The Koonin formula expresses the 
correlation function through the nonrelativistic wave function of two 
particles of interest. Since the observed correlation functions are 
significantly different from unity only for small relative momenta when 
the relative motion of particles is nonrelativistic, it is legitimate 
to use the nonrelativistic wave function in the center-of-mass frame 
of two particles. However, it requires an explicit transformation of 
the source function to the center-of-mass frame. It should be mentioned 
that transformation properties of nonrelativistic wave function 
under a Lorentz boost are not well understood. Only recently it has 
been shown using the Bethe-Salpeter equation that the hydrogen atom 
wave function experiences the Lorentz contraction \cite{Jarvinen:2004pi} 
under the Lorentz boost. Therefore, we perform the calculation in 
the center-of-mass frame of the pair and then we transform the correlation
function to the source rest frame.  

Throughout our whole analysis the source function is of the Gaussian 
form. Such a choice has several advantages. First of all, the Gaussian 
source functions are often used to describe experimental data. Actually, 
the imaging method 
\cite{Brown:1997ku,Brown:2005ze} shows that non-Gaussian contributions 
to the source functions are at a percent level 
\cite{Panitkin:2001qb,Chung:2002vk,Adler:2006as,Afanasiev:2007kk,Alt2008}. 
There are also pure theoretical advantages of the Gaussian parameterization.
When the single particle source function is Gaussian, the so-called 
relative source function is Gaussian as well. Since the Gaussian source 
function has a simple Lorentz covariant form, Lorentz transformations 
can be easily performed. Due to the two features of the Gaussian 
source functions, our calculations are mainly analytical which in 
turn allowed us, in particular, to carefully study relativistic effects
mentioned above. 

The Gaussian parameterization we use has an important disadvantage 
- the fireball expansion is entirely neglected. The study of source 
expansion, however, goes beyond the scope of our analysis. We address 
in this paper a specific question whether the Bowler-Sinyukov procedure 
properly removes Coulomb effects from the correlation functions. For 
this purpose we compute the Coulomb correlation function with the 
Gaussian source, we apply the Bowler-Sinyukov procedure and we check 
how accurately the free correlation function, which is also computed 
with the Gaussian source, is reproduced. It is certainly of interest 
to study how the fireball's expansion influences the Coulomb effects. 
Before that, however, one should systematically analyze to what extend 
the expanding fireball can be represented by a Gaussian source. For this
reason we do not discuss the interplay of Coulomb interaction and 
fireball expansion. Actually, the problem cannot be addressed using 
the computational methods we developed. 

Throughout the paper we use natural units, where $c = \hbar = 1$, and 
our metric convention is $(+,-,-,-)$.


\section{Definition}
\label{sec-def}


The correlation function $C(\mathbf{p}_{1}, \mathbf{p}_{2})$
of two particles with momenta $\mathbf{p}_{1}$ and $\mathbf{p}_{2}$
is defined as
\be
\label{def-corr-fun}
C({\bf p}_1, {\bf p}_2)
=\frac{\frac{dN}{d{\bf p}_1 d{\bf p}_2}}
{\frac{dN}{d{\bf p}_1}\frac{dN}{d{\bf p}_2}} \;,
\ee
where $\frac{dN}{d\mathbf{p}_1 d\mathbf{p}_2}$ and
$\frac{dN}{d\mathbf{p}_1}$ is, respectively, the two- and one-particle 
momentum distribution. The correlation function can be written down 
in a Lorentz covariant form
\be
\label{def-cov}
C({\bf p}_1, {\bf p}_2)=
\frac{E_1 E_2\frac{dN}{d\mathbf{p}_1 d\mathbf{p}_2}}
{E_1\frac{dN}{d\mathbf{p}_1} E_2\frac{dN}{d\mathbf{p}_2}} \;,
\ee
where $E\frac{dN}{d^3\mathbf{p}}$ is the Lorentz invariant
distribution.

The covariant form (\ref{def-cov}) shows that the correlation function 
is a Lorentz scalar field which can be easily transformed from one reference 
frame to another. If the particle four-momenta, which are on mass-shell, 
transform as ${\bf p}_i \rightarrow {\bf p'}_i$ with $i=1,2$, the 
transformed correlation function equals
$$
C'(\mathbf{p}_1'(\mathbf{p}_{1}), \mathbf{p}_2'(\mathbf{p}_{2}))
= C(\mathbf{p}_1, \mathbf{p}_2).
$$

\begin{figure}[t]
\begin{minipage}{5.7cm}
\centering
\includegraphics*[width=5.9cm]{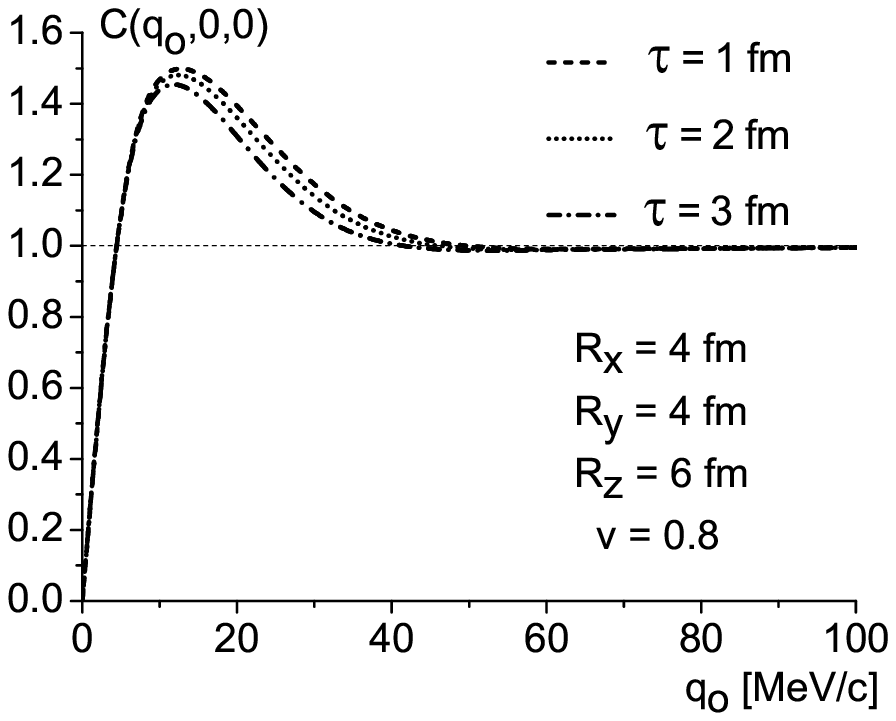}
\vspace{-5mm}
\end{minipage}\hspace{2mm}
\begin{minipage}{5.7cm}
\centering
\includegraphics*[width=5.9cm]{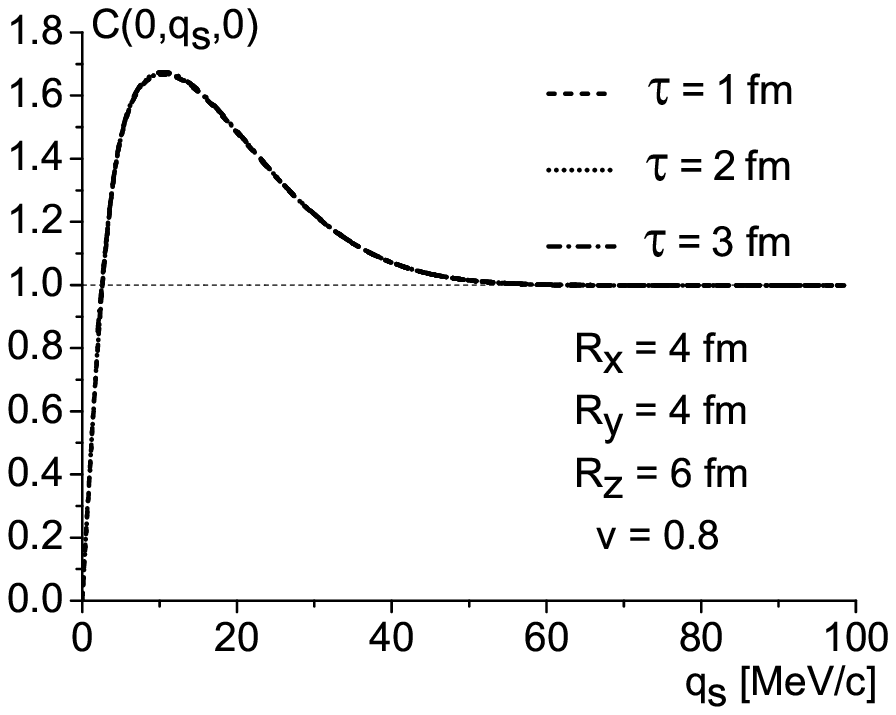}
\end{minipage}\hspace{2mm}
\begin{minipage}{5.7cm}
\centering
\includegraphics*[width=5.9cm]{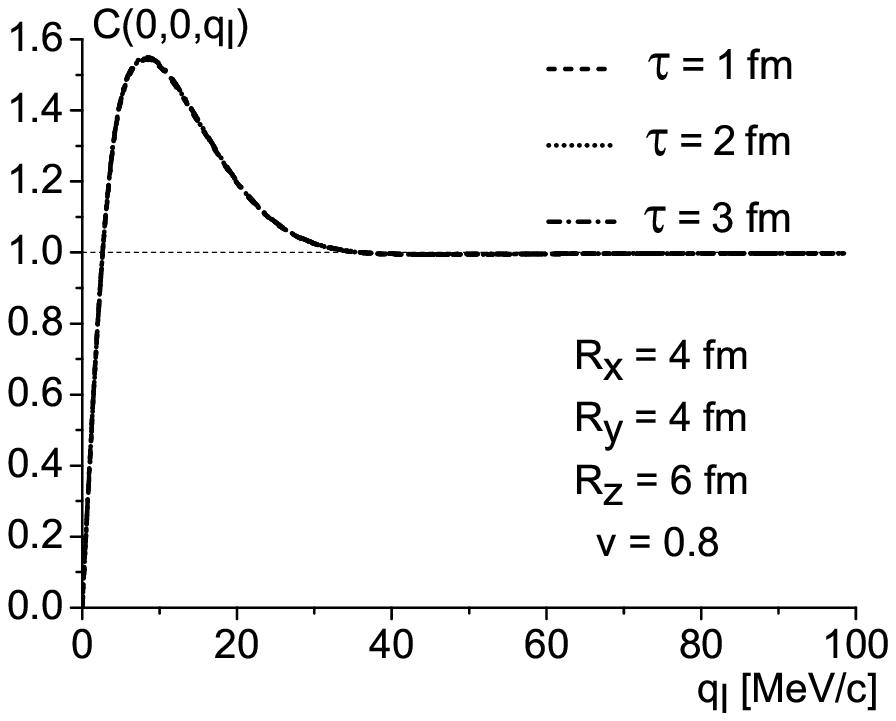}
\end{minipage}
\vspace{-0.3cm}
\caption{The $\pi \pi$ Coulomb correlation functions 
$C(q_o,0,0)$ (left panel), $C(0,q_s,0)$ (central panel) 
and $C(0,0,q_l)$ (right panel) as functions of $q_o$,
$q_s$ or $q_l$, respectively, for three values of the 
emission time $\tau = 1, 2, 3$ fm. The remaining 
parameters are: $R_x = 4$~fm, $R_y = 4$~fm, $R_z = 6$~fm, 
and ${\bf v} = (0.8,0,0)$.}
\label{fig-out-side-long-pi-pi}
\end{figure}


\section{Nonrelativistic Koonin Formula}
\label{sec-koonin}


Within the Koonin model \cite{Koonin:1977fh}, the correlation
function $C$ can be expressed in the source rest frame as
\be 
\label{Koonin} 
C(\mathbf{p}_1,\mathbf{p}_2) = 
\int d^3r_1 dt_1 d^3r_2 dt_2 D(t_1,\mathbf{r}_1) \:
D(t_2,\mathbf{r}_2) \: |\Psi(\mathbf{r}'_1,\mathbf{r}'_2)|^2 \;, 
\ee
where $\mathbf{r}'_i \equiv \mathbf{r}_i+\mathbf{v}_i t_i$,
$\Psi(\mathbf{r}'_1,\mathbf{r}'_2)$ is the wave function of 
the two particles and $D(t,\mathbf{r})$ is the single-particle 
source function which gives the probability to emit the particle 
from the space-time point $(t,\mathbf{r})$. The source function
is normalized as 
\be
\label{norma}
\int d^3r \, dt \, D(t,\mathbf{r})=1 \;.
\ee

After changing the variables ${\bf r}' \leftrightarrow {\bf r}$, the 
correlation function can be written in the form
$$
C(\mathbf{p}_{1}, \mathbf{p}_{2}) = 
\int d^3 r_1 dt_1 d^3 r_2 dt_2 
D(t_1,\mathbf{r}_1-\mathbf{v}_1t_1) \:
D(t_2,\mathbf{r}_2-\mathbf{v}_2t_2)
|\Psi(\mathbf{r}_1,\mathbf{r}_2)|^2 \;.
$$

Now, we introduce the center-of-mass coordinates
\ban
\begin{array}{cc}
\mathbf{r}=\mathbf{r}_2-\mathbf{r}_1, &
\mathbf{R}=\frac{1}{M}(m_1\mathbf{r}_1+m_2\mathbf{r}_2), 
\\[2mm]
t=t_2-t_1, & T=\frac{1}{M}(m_1 t_1+m_2 t_2),
\\[2mm]
\mathbf{q}= \frac{1}{M}(m_2 \mathbf{p}_1 - m_1\mathbf{p}_2),&
\mathbf{P}=\mathbf{p}_1+\mathbf{p}_2,
\end{array}
\ean
where $M \equiv m_1+m_2$. Using the center-of-mass
variables, one gets
\be 
\label{Koonin-r} 
C(\mathbf{q})= 
\int d^3r\: D_r(\mathbf{r})
|\varphi_\mathbf{q}(\mathbf{r})|^2, 
\ee
where the `effective relative' source function is defined as
\be
D_r(\mathbf{r})\equiv \int dt\: D_r(t,\mathbf{r-v}t) \;,
\ee
and the `relative' source function is expressed through
the single-particle source function in the following way
\be
\label{source-relat}
D_r(t,\mathbf{r}) \equiv \int d^3R \, dT \:
D(T-\frac{m_2}{M}t,\mathbf{R}-\frac{m_2}{M}\mathbf{r}) \:
D(T+\frac{m_1}{M}t,\mathbf{R}+\frac{m_1}{M}\mathbf{r}) \,.
\ee
We note that due to the normalization (\ref{norma}), the
functions $D_r(\mathbf{r})$ and $D_r(\mathbf{r-v}t,t)$
are also normalized
\be
\int d^3r \: D_r(\mathbf{r}) = \int d^3r \: dt\: 
D_r(t,\mathbf{r}) = 1 \,.
\ee
To get Eq.~(\ref{Koonin-r}), the wave function was factorized as
$$
\Psi(\mathbf{r}_1,\mathbf{r}_2)=
e^{i\mathbf{P}\mathbf{R}}\varphi_\mathbf{q}(\mathbf{r})
$$
with $\varphi_\mathbf{q}(\mathbf{r})$ being the wave function of 
the relative motion in the center-of-mass frame. Deriving
Eq.~(\ref{Koonin-r}), it has been assumed that the particle velocity,
which enters the effective source function, is the same for both
particles. Thus, we have assumed that 
$\mathbf{v}_1=\mathbf{v}_2=\mathbf{v}$ which
requires, strictly speaking, ${\bf q} = 0$. However, one observes
that $|{\bf v}_1 - {\bf v}_2 | \ll |{\bf v}_i|$ if 
$|{\bf q} | \ll \mu |{\bf p}_i|/m_i$ where $\mu \equiv m_1 m_2/M$. 
Thus, the approximation $\mathbf{v}_1 \approx \mathbf{v}_2$ holds for  
sufficiently small particle's momenta in the center-of-mass frame. 
It should be stressed that the dependence of the correlation function 
on ${\bf q}$ is mostly controlled by the dependence of the wave 
function on ${\bf q}$ which is not influenced by  the above approximation.

\begin{figure}[t]
\begin{minipage}{5.7cm}
\centering
\includegraphics*[width=5.9cm]{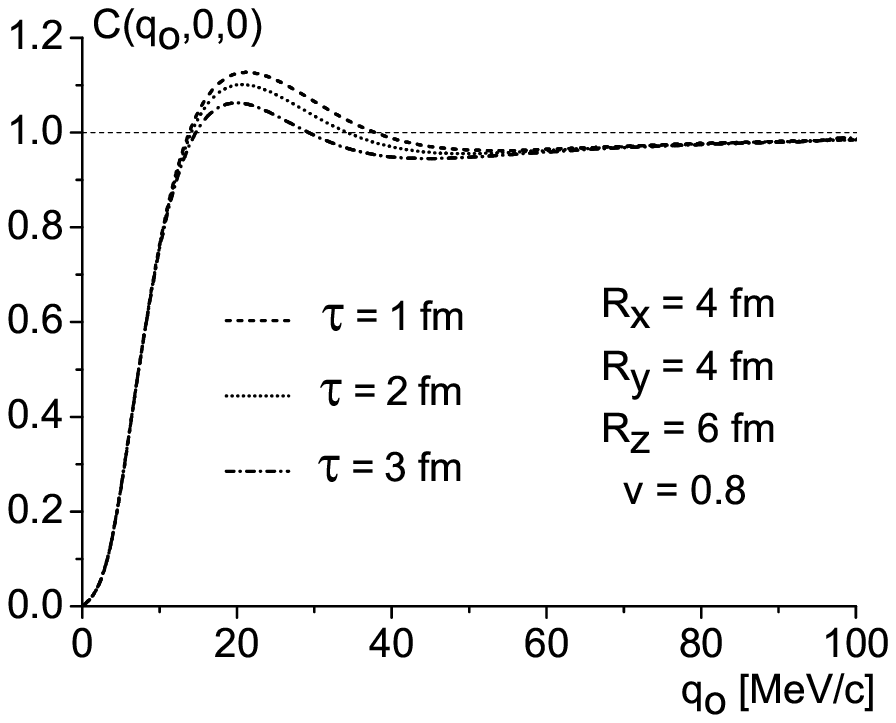}
\end{minipage}\hspace{2mm}
\centering
\begin{minipage}{5.7cm}
\includegraphics*[width=5.9cm]{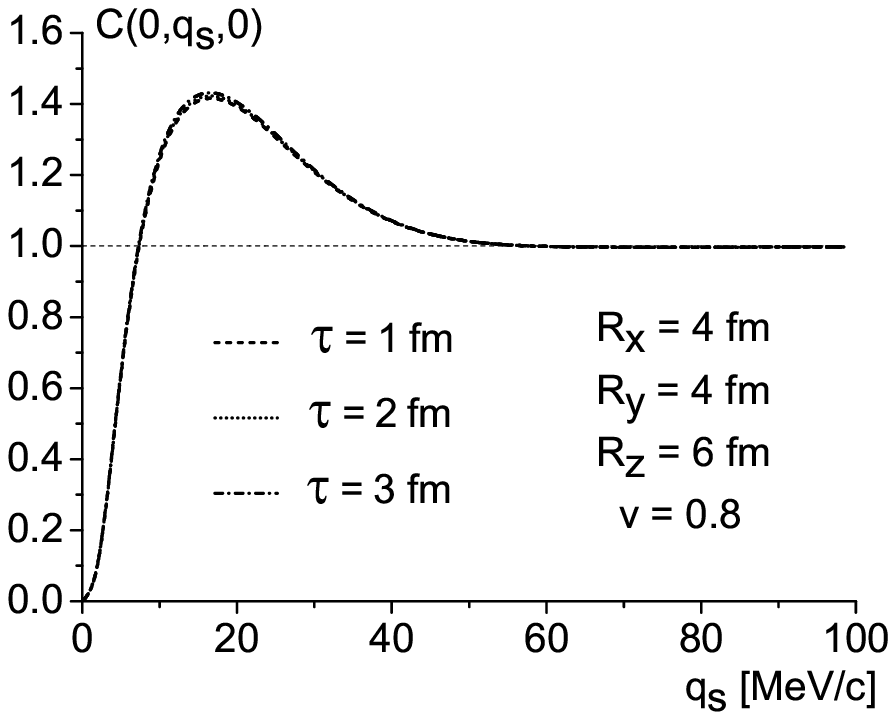}
\end{minipage}\hspace{2mm}
\centering
\begin{minipage}{5.7cm}
\includegraphics*[width=5.9cm]{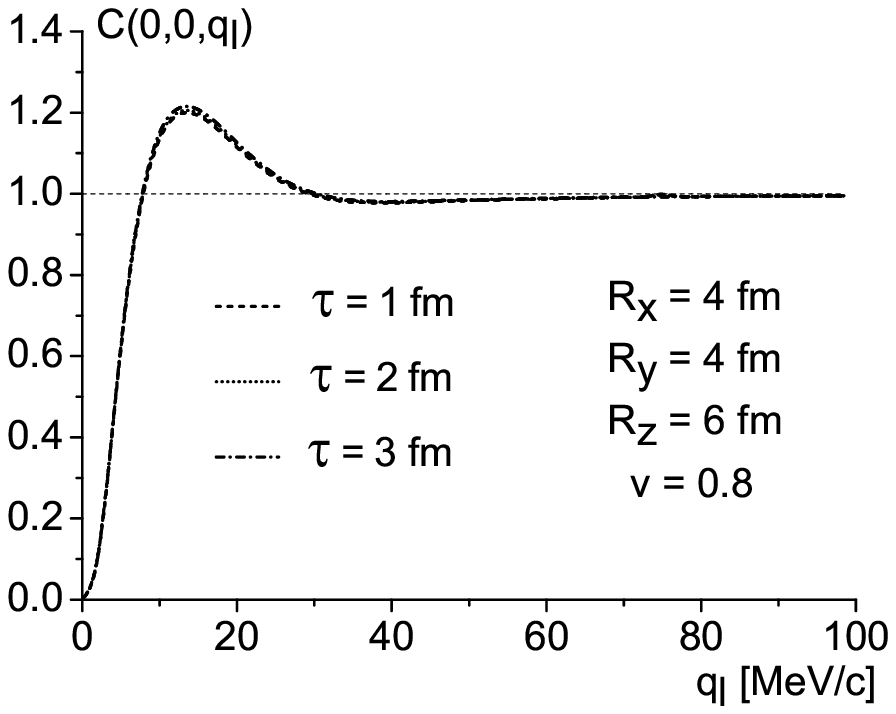}
\end{minipage}
\vspace{-0.3cm}
\caption{The $KK$ Coulomb correlation functions $C(q_o,0,0)$ 
(left panel), $C(0,q_s,0)$ (central panel) and $C(0,0,q_l)$
(right panel) as functions of $q_o$, $q_s$ or $q_l$, 
respectively, for three values of the 
emission time $\tau = 1, 2, 3$ fm. The remaining 
parameters are: $R_x = 4$~fm, $R_y = 4$~fm, $R_z = 6$~fm, 
and ${\bf v} = (0.8,0,0)$.}
\label{fig-out-side-long-K-K}
\end{figure}

We choose the Gaussian form of the single-particle source function
$D(t,{\bf r})$ 
\be
\label{gauss}
D(t,\mathbf{r})=\frac{1}{4\pi^2 R_x R_y R_z \tau} \:
\exp\Big[- \frac{t^2}{2\tau^2} -\frac{x^2}{2R_x^2} 
- \frac{y^2}{2R_y^2} - \frac{z^2}{2R_z^2}\Big],
\ee
where ${\bf r}=(x,y,z)$ and the parameters $\tau$, $R_x$, $R_y$
and $R_z$ characterize the lifetime and sizes of the source.
Specifically, the parameters $\tau$ and $R_x$ give, respectively,
$$
\tau^2 = \langle t^2 \rangle \equiv \int d^3r\:dt\: t^2
D(t,\mathbf{r})
\;,\;\;\;\;\;\;\;\;\;
R_x^2 = \langle x^2\rangle \equiv \int d^3r \: dt\: x^2
D(t,\mathbf{r})\;.
$$

The relative source function computed from Eq.~(\ref{source-relat})
with the single-particle source (\ref{gauss}) is
\be
\label{gauss-relat}
D_r(t,\mathbf{r})=\frac{1}{16\pi^2 R_x R_y R_z \tau} \:
\exp\Big[ -\frac{t ^2}{4\tau^2} -\frac{x^2}{4R_x^2}
-\frac{y^2}{4R_y^2}-\frac{z^2}{4R_z^2}\Big].
\ee
We note that the particle's masses, which are present in the definition
(\ref{source-relat}), disappear completely in the formula (\ref{gauss-relat}).
This is the feature of the Gaussian parameterization (\ref{gauss}). 

In the case of non-interacting identical bosons, the two-particle
symmetrized wave function is
$$
\Psi(\mathbf{r}_1,\mathbf{r}_2)= \frac{1}{\sqrt{2}}[e^{i\mathbf{p}_1\mathbf{r}_1+\mathbf{p}_2\mathbf{r}_2}
+ e^{i\mathbf{p}_2\mathbf{r}_1+\mathbf{p}_1\mathbf{r}_2}]=
\frac{1}{\sqrt{2}}[e^{i\mathbf{q}\mathbf{r}}+
 e^{-i\mathbf{q}\mathbf{r}}]e^{i\mathbf{PR}}.
$$
It gives the modulus square of the wave function of relative motion 
$|\varphi_{\bf q}({\bf r})|^2=1+\cos{(2{\bf q}{\bf r})}$ which in turn
provides the correlation function equal to
\be 
\label{corr-free}
C({\bf q}) = 1 + \exp \big[ - 4 \big(
\tau^2({\bf qv})^2 + R_x^2 q_x^2 + R_y^2 q_y^2 + R_z^2 q_z^2 \big)
\big] \;,
\ee
where ${\bf q} \equiv (q_x,q_y,q_z)$. We note that the `cross terms' such 
as $q_x q_z$, which are discussed in \cite{Chapman:1994yv}, do not show 
up, as the source function (\ref{gauss}) obeys the mirror symmetry 
$D(t,{\bf r}) = D(-t,-{\bf r})$. We also note that ${\bf q}$ often 
denotes the relative momentum ${\bf p}_1 - {\bf p}_2$ not the momentum 
in the center-of-mass frame, which for equal mass nonrelativistic particles
equals $\frac{1}{2}({\bf p}_1 - {\bf p}_2)$, and then, the factor 4 does 
not show up in the correlation function (\ref{corr-free}) of identical 
free bosons. However, we believe that using the momentum in the 
center-of-mass frame is physically better motivated, as the center-of-mass
variables naturally appear when the center-of-mass motion is separated 
from the relative one.


\section{Relativistic Formulations}
\label{sec-relat}


There are two natural ways to `relativize' the Koonin formula (\ref{Koonin}).
The first one provides an explicitly Lorentz covariant correlation function but
it is applicable only for the non-interacting particles. The second one holds
only in a specific reference frame but it is applicable for interacting 
particles as well. Below, we consider the two methods.  We start, however, 
with the discussion of the Lorentz covariant form of the source function.

\begin{figure}[t]
\begin{minipage}{5.7cm}
\centering
\includegraphics*[width=5.9cm]{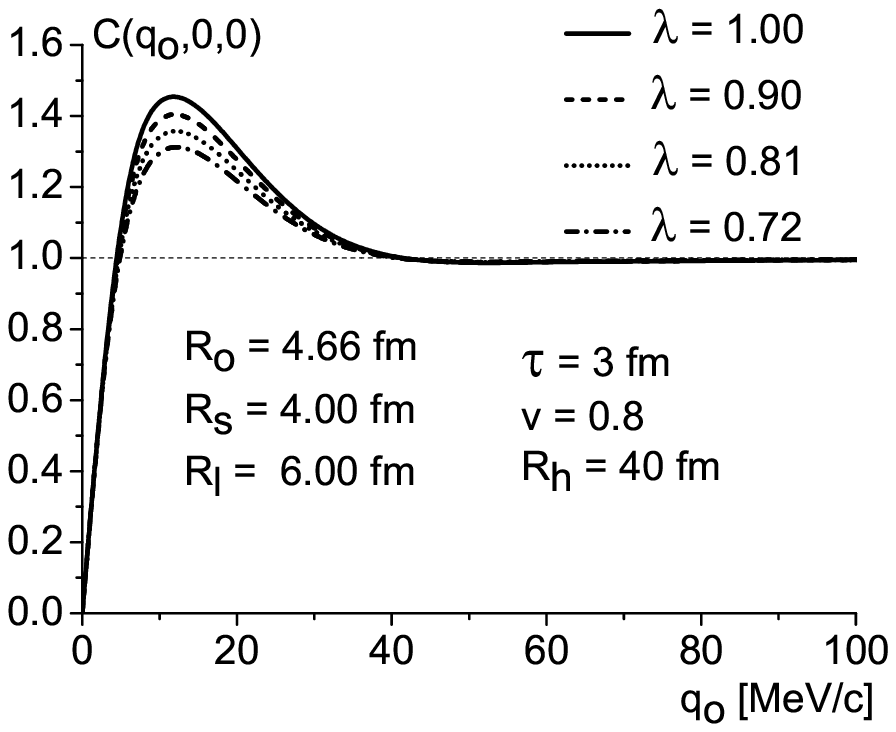}
\end{minipage}\hspace{2mm}
\begin{minipage}{5.7cm}
\centering
\includegraphics*[width=5.9cm]{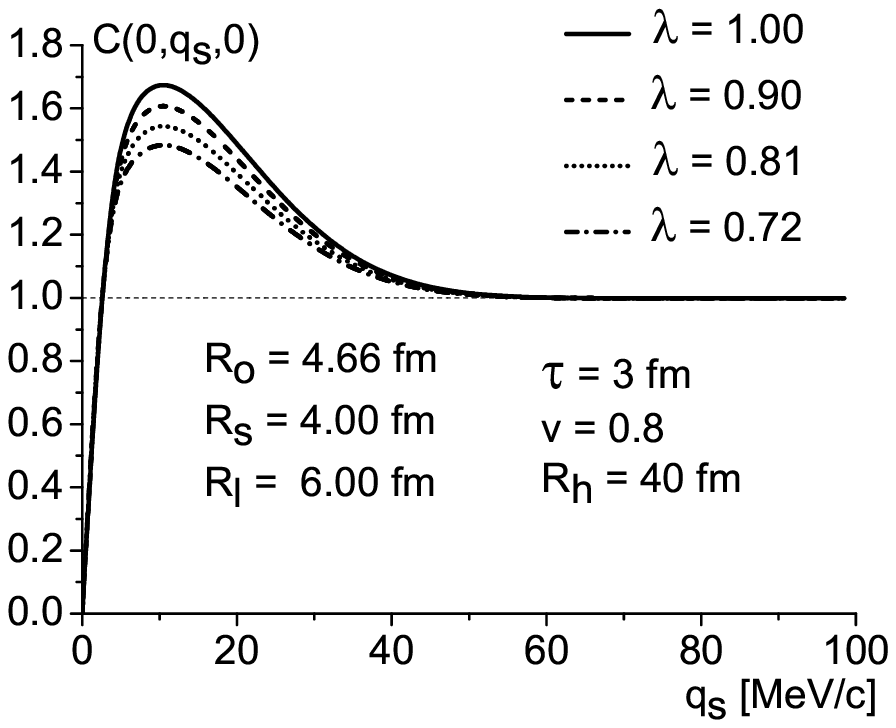}
\end{minipage}\hspace{2mm}
\begin{minipage}{5.7cm}
\centering
\includegraphics*[width=5.9cm]{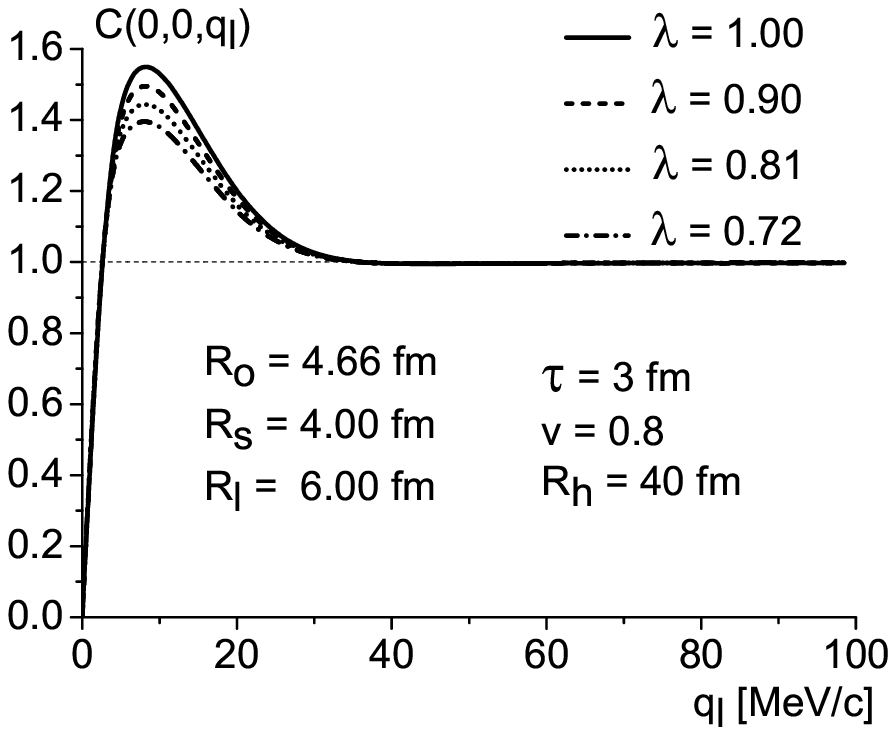}
\end{minipage}
\vspace{-0.3cm}
\caption{The $\pi \pi$ Coulomb correlation functions 
$C(q_o,0,0)$ (left panel), $C(0,q_s,0)$ (central panel) 
and  $C(0,0,q_l)$ (right panel) as functions of $q_o$,
$q_s$ or $q_l$, respectively, for various halo contributions.
The halo is spherical with $R_h = 40$~fm and $\tau_h =0$. The
fireball parameters are $R_x = 4$~fm, $R_y = 4$~fm, $R_z = 6$~fm, 
$\tau=3$~fm. The pair velocity is ${\bf v} = (0.8,0,0)$.}
\label{fig-halo-lam}
\end{figure}

\subsection{Lorentz covariant source function}

Because of its probabilistic interpretation, the source function
transforms under Lorentz transformation as a scalar field. Therefore,
the covariant form of the Gaussian parameterization of the source 
function (\ref{gauss}) is written as \cite{Heinz:1996qu}
\be
\label{source-cov}
D(x)=\frac{\sqrt{{\rm det}\Lambda}}{4\pi^2} \;
{\rm exp} [-\frac{1}{2}x_\mu \Lambda^{\mu\nu}x_\nu],
\ee
where $x^\mu$ is the position four-vector and $\Lambda^{\mu\nu}$
is the Lorentz tensor characterizing the source which in the
source rest frame is
\be
\label{source-matrix}
\Lambda^{\mu\nu}=\left[\begin{array}{cccc}
\frac{1}{\tau^2} & 0 & 0 & 0  \\
0 & \frac{1}{R_{x}^2} & 0 & 0  \\
0 & 0 & \frac{1}{R_{y}^2} & 0  \\
0 & 0 & 0 & \frac{1}{R_{z}^2}  \\
\end{array}\right].
\ee
The source function as written in Eq.~(\ref{source-cov}) obeys
the normalization condition (\ref{norma}) not only for the diagonal
matrix $\Lambda$ but for non-diagonal as well. 

The source function (\ref{source-cov}) is evidently the Lorentz 
scalar that is
\be
D'(x') = \frac{\sqrt{{\rm det} \Lambda'}}{4 \pi^2}
\exp{[-\frac{1}{2} x'_\mu \Lambda'^{\mu\nu}x'_\nu]}
= \frac{\sqrt{{\rm det} \Lambda}}{4 \pi^2}
\exp{[-\frac{1}{2} x_\mu \Lambda^{\mu\nu}x_\nu]}
=D(x) \;,
\ee
where $x'_\mu = L_{\mu}^{\;\;\nu}x_\nu$
and $\Lambda'^{\mu\nu} = L_{\;\;\sigma}^{\mu}\Lambda^{\sigma\rho}L_{\rho}^{\;\;\nu}$
with $L_{\sigma}^{\mu}$ being the matrix of Lorentz transformation. 
We note that ${\rm det} \Lambda' = {\rm det}L \: {\rm det} \Lambda
\: {\rm det}L^{-1} = {\rm det} \Lambda$. 

The covariant relative source function (\ref{gauss-relat}) 
is given by
\be
\label{source-rel-cov}
D_r(x) = \frac{\sqrt{{\rm det} \Lambda}}{16\pi^2}
\exp{[-\frac{1}{4} x_\mu \Lambda^{\mu\nu}x_\nu]}\;. 
\ee

\begin{figure}[t]
\begin{minipage}{8cm}
\centering
\includegraphics*[width=6.5cm]{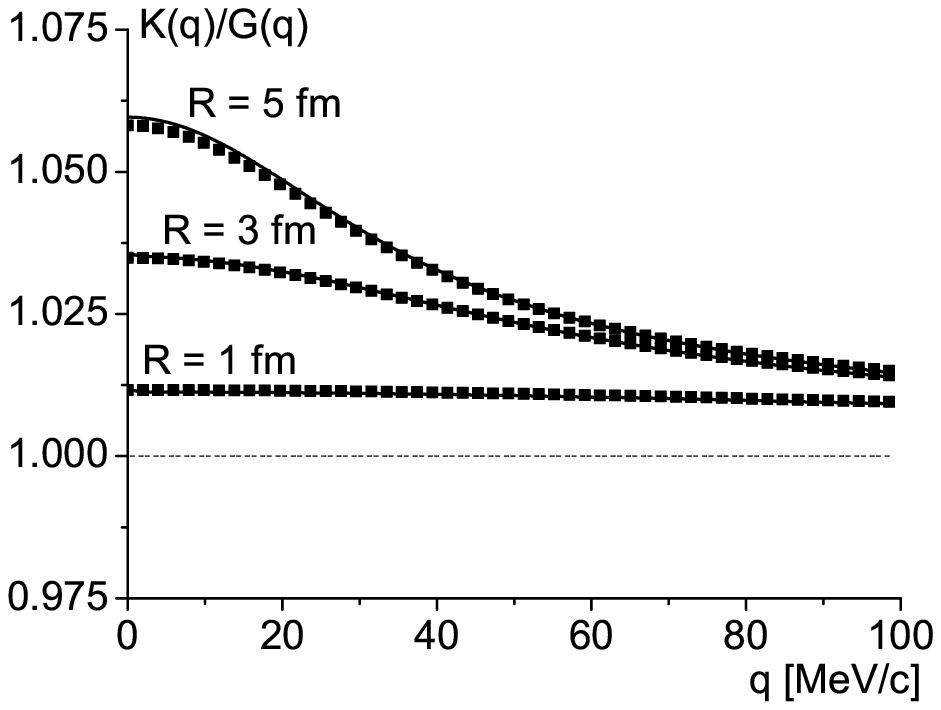}
\end{minipage}\hspace{3mm}
\centering
\begin{minipage}{8cm}
\includegraphics*[width=6.5cm]{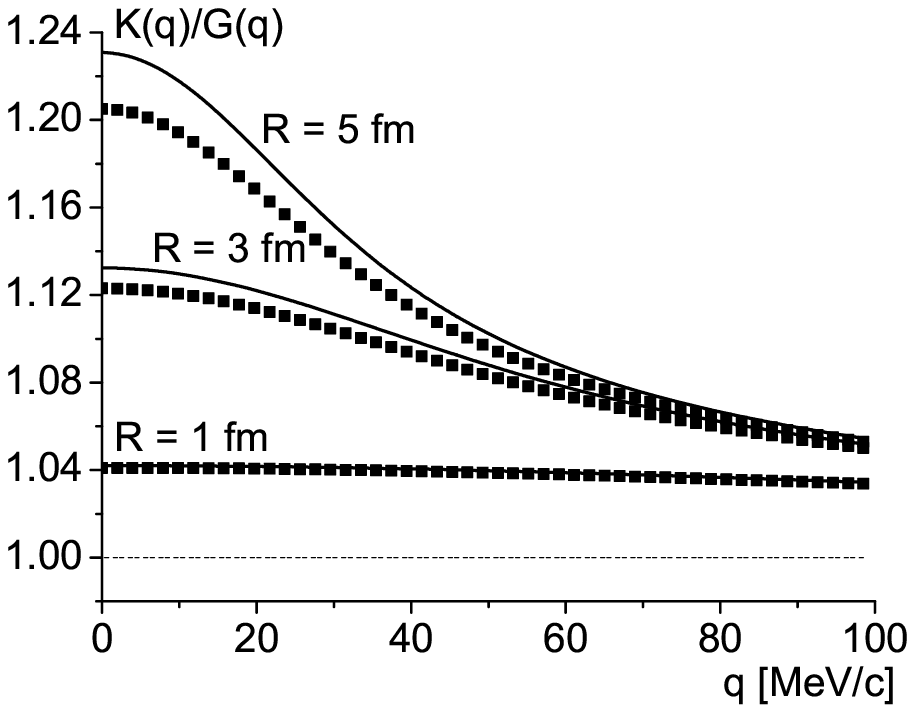}
\end{minipage}
\caption{The correction factor $K(q)$ divided by the Gamov 
factor $G(q)$ as function of $q$ for source radii $R=1,\,3,\,5$
fm. The left panel is for pions and the right one for kaons.
The solid lines and squares represent, respectively, the 
exact formula (\ref{K-exact}) and the approximate one 
(\ref{B-S-pop-appr}).}
\label{popandapro}
\end{figure}

\subsection{Explicitly covariant `relativization'}
\label{sub-sec-exp-cov}

As follows from Eq.~(\ref{def-cov}), the correlation function is
a Lorentz scalar. Therefore, the Koonin formula (\ref{Koonin}) can 
be `relativized' demanding its Lorentz covariance. Let us write the 
formula as
\be
\label{explicit}
C(p_1,p_2) = \int d^4x_1d^4x_2 D(x_1) \:
D(x_2)|\Psi(x_1,x_2)|^2,
\ee
where $p_i$ and $x_i$ is, respectively, the four-momentum and
four-position. Since the source function $D(x)$ and the four-volume
element $d^4x_i$ are both the Lorentz scalars, the whole formula 
(\ref{explicit}) is covariant if the wave function $\Psi(x_1,x_2)$ 
is covariant as well. In the case of non-interacting bosons the 
relativistic wave function $\Psi(x_1,x_2)$
is
\be
\label{rel-wave-fun}
\Psi(x_1,x_2)= \frac{1}{\sqrt{2}}
(e^{ip_1 x_1 + i p_2 x_2} + e^{ip_1 x_2 + i p_2 x_1}).
\ee
As the function depends on the scalar products of two four-vectors,
it is the Lorentz scalar. We note that the function 
(\ref{rel-wave-fun}) depends on two time arguments. 

Our further considerations are limited to pairs of identical 
particles and thus, we introduce the relative coordinates: 
\ba
\label{relative}
\begin{array}{cc}
x = x_2-x_1, & X = \frac{1}{2}(x_1 + x_2), 
\\[2mm]
q = \frac{1}{2}(p_1 - p_2),& P = p_1 + p_2.
\end{array}
\ea
In this section and in Appendix A $q$ is the four-vector $(q_0, {\bf q})$ 
but in the remaining sections $q \equiv |{\bf q}|$. Hopefully, it will not
cause any confusion.

We note that in the nonrelativistic treatment the three-vectors
${\bf r}$ and ${\bf q}$, which are given by the four-vectors 
$x=(t,{\bf r})$ and $q=(q_0, {\bf q})$, correspond to the
inter-particle separation and the particle's momentum in the 
center-of-mass of the pair. This is, however, not the case in 
the relativistic domain. To get the center-of-mass 
variables, the four-vectors need to be Lorentz transformed. 
We also note that $q_0 = {\bf qv}$ which is proven as
\footnote{We are very grateful to Richard Lednicky for calling
our attention to the fact that $q_0 = {\bf qv}$ not only for small 
${\bf q}$, as we erroneously thought, but for any ${\bf q}$.}
$$
q_0 \equiv \frac{1}{2} \Big(\sqrt{m^2 +{\bf p}_1^2} 
- \sqrt{m^2 +{\bf p}_2^2} \Big)
= \frac{1}{2} \frac{{\bf p}_1^2 - {\bf p}_2^2}
{\sqrt{m^2 +{\bf p}_1^2} + \sqrt{m^2 +{\bf p}_2^2}}
= {\bf q} \frac{{\bf p}_1 + {\bf p}_2}
{\sqrt{m^2 +{\bf p}_1^2} + \sqrt{m^2 +{\bf p}_2^2}} 
= {\bf qv}\;.
$$

With the variables (\ref{relative}), the wave function 
(\ref{rel-wave-fun}) equals
$$
\Psi(x,X)=\frac{1}{\sqrt{2}}(e^{iqx}+e^{-iqx})e^{iPX} \;,
$$
and the correlation function is found in the form
$$
C(q)=1+\exp[-4q_\mu(\Lambda^{\mu\nu})^{-1}q_\nu] \;,
$$
which is explicitly Lorentz covariant. For the source matrix 
(\ref{source-matrix}), the correlation function equals
\be
\label{corr-free-relat}
C(q)= 1+ \exp 
\big[-4(q_0^2\tau^2 + q_x^2R_x^2 + q_y^2R_y^2 + q_z^2R_z^2)\big] \;.
\ee
Since $q_0 = {\bf qv}$, the correlation function 
(\ref{corr-free-relat}) exactly coincides with the nonrelativistic 
expression (\ref{corr-free}). This coincidence is not completely
obvious as the time variables enter differently in the Koonin
formula (\ref{Koonin}) and in the covariant one (\ref{explicit}).

\begin{figure}[t]
\begin{minipage}{5.7cm}
\centering
\includegraphics*[width=5.9cm]{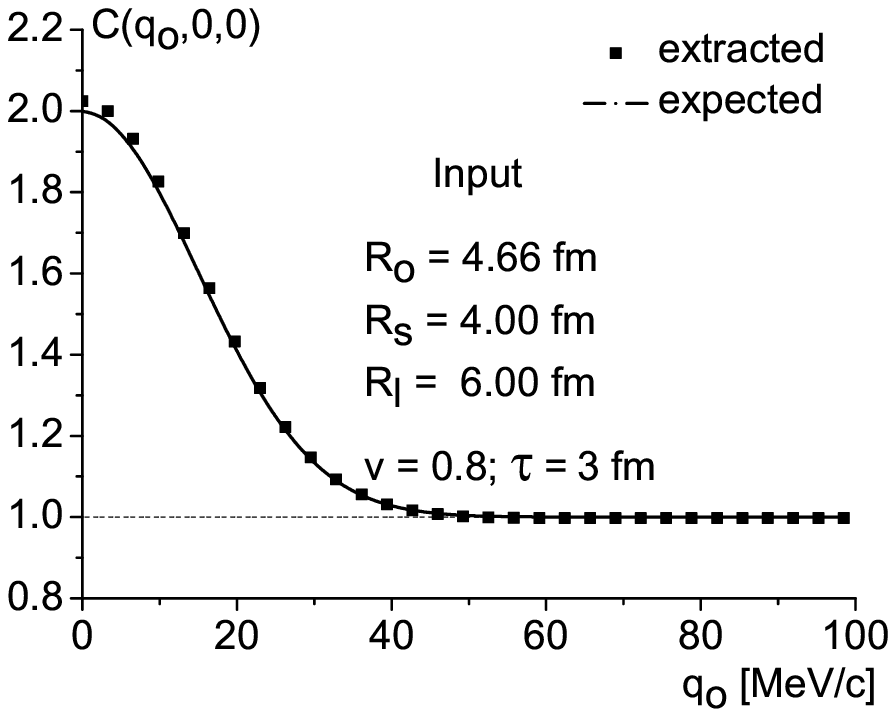}
\end{minipage}\hspace{2mm}
\begin{minipage}{5.7cm}
\centering
\includegraphics*[width=5.9cm]{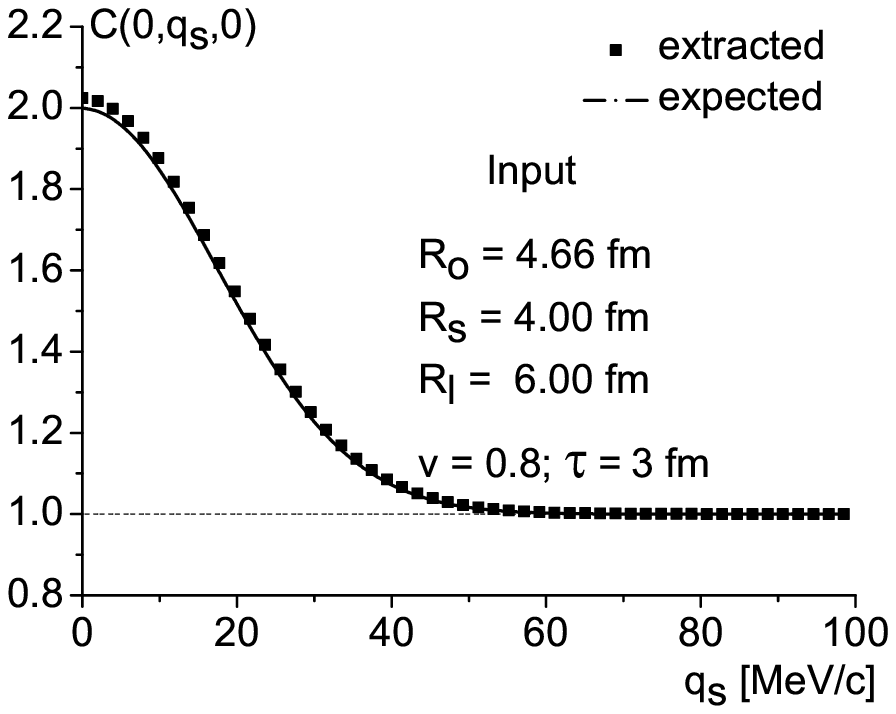}
\end{minipage}\hspace{2mm}
\begin{minipage}{5.7cm}
\centering
\includegraphics*[width=5.9cm]{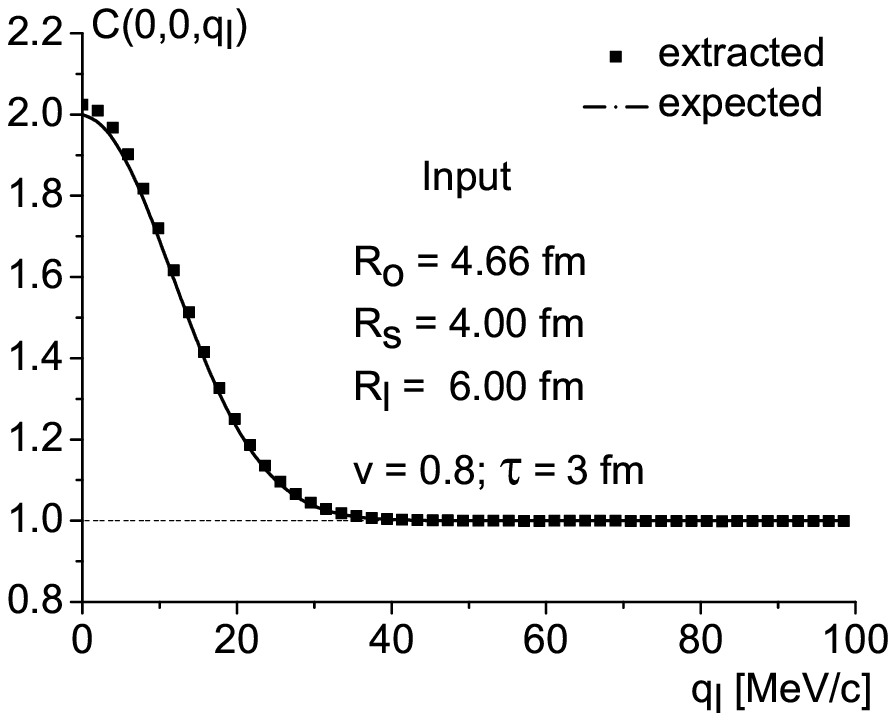}
\end{minipage}
\vspace{-0.3cm}
\caption{The $\pi \pi$ free correlation functions 
$C(q_o,0,0)$ (left panel), $C(0,q_s,0)$ (central panel) 
and  $C(0,0,q_l)$ (right panel) as a function of $q_o$,
$q_s$ or $q_l$, respectively. The fireball parameters are 
$R_x = 4$~fm, $R_y = 4$~fm, $R_z = 6$~fm, $\tau=3$~fm and 
the pair velocity is ${\bf v} = (0.8,0,0)$. The extracted 
`free' functions are represented by the squares and the 
expected free correlation functions by the solid lines.}
\label{fig-free}
\end{figure}

Let us consider the correlation function in the center-of-mass
frame of the particle pair. We assume that the velocity of 
the center-of-mass frame in the source rest frame is along the 
axis $x$. Then,  ${\bf v}=(v,0,0)$ and $q_0=q_xv$. 
The correlation function (\ref{corr-free-relat}), which holds in 
the source rest frame, equals
\be 
\label{corr-free-relat-vx}
C(\mathbf{q})=1+ \exp
\big[-4\big( (v^2\tau^2 + R_x^2)q_x^2 
+ R_y^2 q_y^2+ R_z^2 q_z^2 \big)\big]\;.
\ee
As seen, the effective source radius in the direction $x$ is
$\sqrt{R_x^2 + v^2\tau^2}$. We now transform the source function 
to the center-of-mass frame where the quantities are labeled with 
the index $*$. The center-of-mass source matrix (\ref{source-matrix}), 
which is computed as 
$$
\Lambda^{\mu\nu}_* = L_{\;\;\sigma}^{\mu}
\Lambda^{\sigma\rho} L_{\rho}^{\;\;\nu},
$$
where
$$
L_{\;\;\sigma}^{\mu}=\left[\begin{array}{cccc}
\gamma & -v\gamma & 0 & 0  \\
-v\gamma & \gamma & 0 & 0  \\
0 & 0 & 1 & 0  \\
0 & 0 & 0 & 1  \\
\end{array}\right] 
$$
with $\gamma \equiv (1-v^2)^{-1/2}$, equals
\ba
\label{source-matrix-CM}
\Lambda^{\mu\nu}_*=\left[\begin{array}{cccc}
\gamma^2(\frac{1}{\tau^2}+\frac{v^2}{R_{x}^2}) & 
- \gamma^2v(\frac{1}{\tau^2}+\frac{1}{R_{x}^2}) & 0 & 0  \\
-\gamma^2v(\frac{1}{\tau^2}+\frac{1}{R_{x}^2}) & 
\gamma^2(\frac{v^2}{\tau^2}+\frac{1}{R_{x}^2}) & 0 & 0  \\
0 & 0 & \frac{1}{R_{y}^2} & 0  \\
0 & 0 & 0 & \frac{1}{R_{z}^2}  \\
\end{array}\right] \;.
\ea
Then, the correlation function in the center-of-mass frame is found 
to be
\be 
\label{corr-free-relat-vx-CM} 
C(q_*) 
= 1 + 
\exp [-4q_{*\mu}(\Lambda^{\mu\nu}_*)^{-1}q_{*\nu} ]
= 1 + \exp \big[-4 \big( \gamma^2(v^2\tau^2 + R_x^2)q_{*x}^2 
+ R_y^2 q_{*y}^2+ R_z^2 q_{*z}^2\big)\big]\;.
\ee 
As seen, the effective source radius along the direction of
the velocity is elongated, not contracted as one can naively
expect, by the factor $\gamma$. 

\begin{figure}[t]
\begin{minipage}{5.7cm}
\centering
\includegraphics*[width=5.7cm]{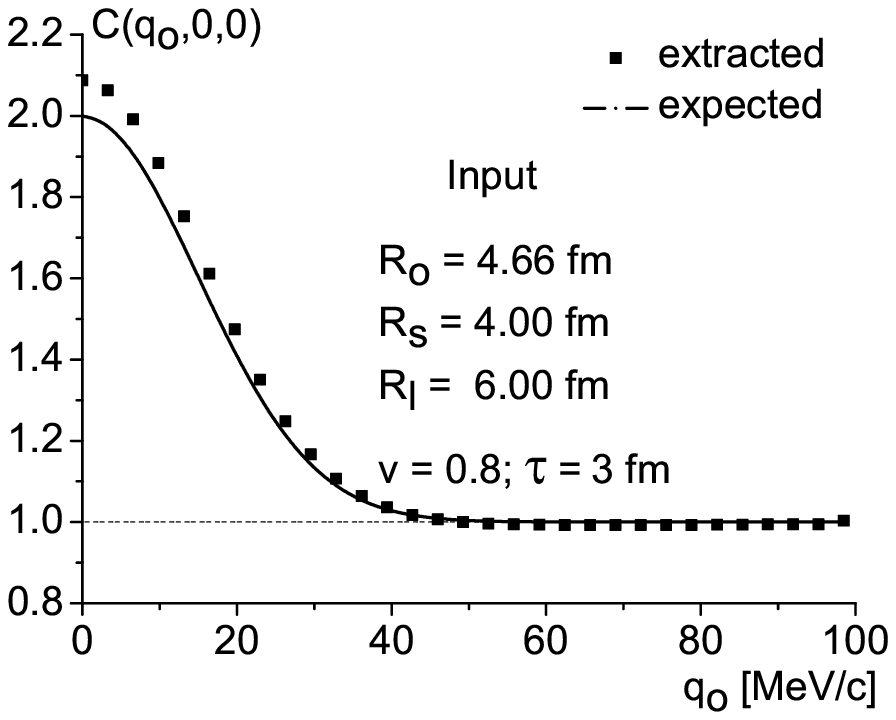}
\end{minipage}\hspace{2mm}
\begin{minipage}{5.7cm}
\centering
\includegraphics*[width=5.7cm]{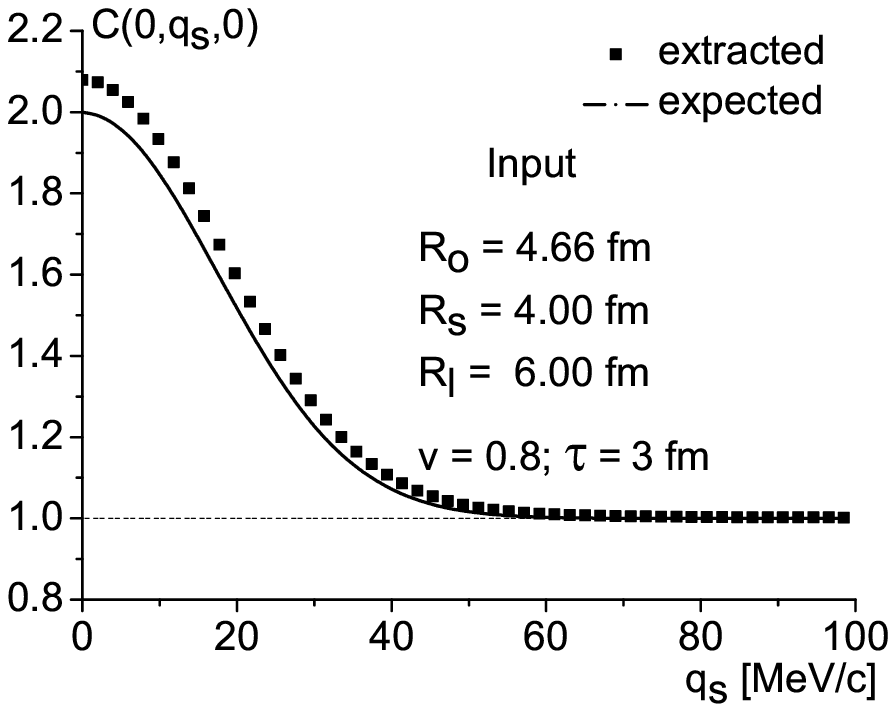}
\end{minipage} \hspace{2mm}
\begin{minipage}{5.7cm}
\centering
\includegraphics*[width=5.7cm]{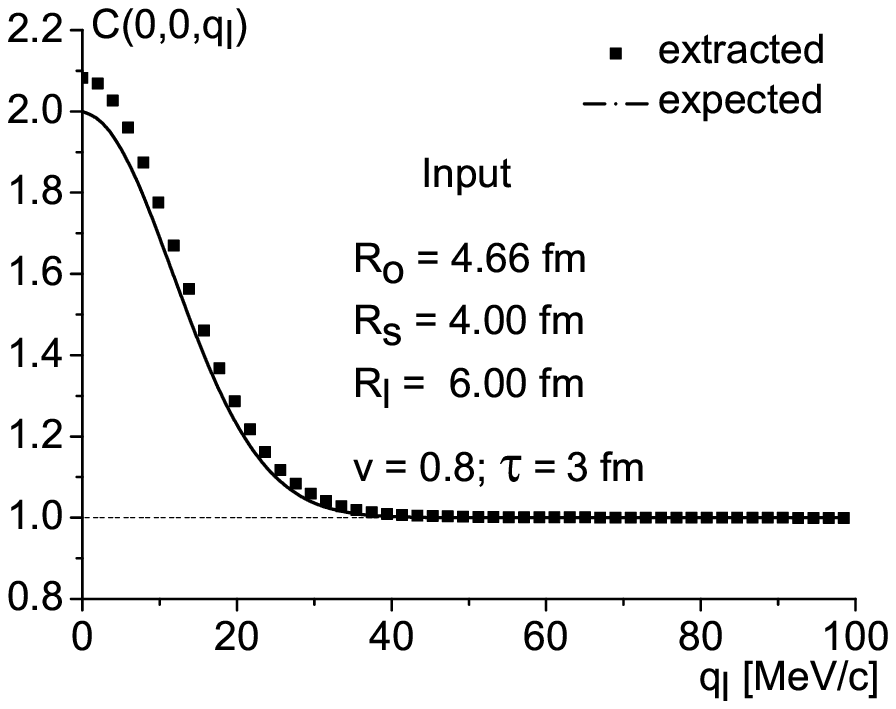}
\end{minipage}
\caption{The $KK$ free correlation function 
$C(q_o,0,0)$ (left panel), $C(0,q_s,0)$ (central panel) 
and  $C(0,0,q_l)$ (right panel) as a function of $q_o$,
$q_s$ or $q_l$, respectively. The fireball parameters are 
$R_x = 4$~fm, $R_y = 4$~fm, $R_z = 6$~fm, $\tau=3$~fm and 
the pair velocity is ${\bf v} = (0.8,0,0)$. The extracted 
`free' functions are represented by the squares and the 
expected free correlation functions by the solid lines.}
\label{fig-free-KK}
\end{figure}

\subsection{Non-covariant relativization}

The quantum mechanical description of two relativistic interacting
particles faces serious difficulties. The problem is greatly
simplified when the relative motion of two particles is
nonrelativistic (with the center-of-mass motion being fully
relativistic). Since the correlation functions usually differ from
unity only for small relative momenta of particles, it is reasonable
to assume that the relative motion is nonrelativistic. We further 
discuss the correlation functions taking into account the
relativistic effects of motion of particles with respect to the
source but the particle's relative motion is treated
nonrelativistically. In such a case, the wave function of relative
motion is a solution of the nonrelativistic Schr\"odinger equation.
Thus, we compute the correlation function directly from the Koonin
formula (\ref{Koonin}) but the computation is performed in the 
center-of-mass frame of the pair. For this reason we first 
transform the source function to this frame and then, after 
performing the integrations over $x_1$ and $x_2$, we transform 
the whole correlation function to the source rest frame.
We stress here that according to the definition (\ref{def-cov}) 
the correlation function is the Lorentz scalar. 

As already noted, we compute the correlation function in the 
center-of-mass frame of the pair using the relative variables 
(\ref{relative}). The correlation function thus equals
\be
\label{Koonin-relat}
C({\bf q}_*) = \int
d^3r_* dt_* \: D_r(t_*,{\bf r}_*) \:
|\varphi_{{\bf q}_*}({\bf r}_*)|^2,
\ee
where $D_r(t_*,{\bf r}_*)$ is the relative source function 
(\ref{source-rel-cov}) and $\varphi_{{\bf q}_*}({\bf r}_*)$ is,
as previously, the nonrelativistic wave function of relative motion. 
The note here that ${\bf v}_* = 0$ by definition. The formula 
(\ref{Koonin-relat}) can be rewritten as 
\be
\label{Koonin-relat-two}
C({\bf q}_*) = \int
d^3 r_* \: D_r({\bf r}_*) \:
|\varphi_{{\bf q}_*}({\bf r}_*)|^2,
\ee
where
\be
\label{source-CM}
D_r({\bf r}_*) \equiv \int dt_* D_r(t_*,{\bf r}_*) =
\frac{1}{8\pi^{3/2}\sqrt{\gamma^2(R_x^2+v^2\tau^2)}R_yR_z}
\exp \left[-\frac{1}{4}
\left(\frac{x_*^2}{\gamma^2(R_x^2+v^2\tau^2)}
+\frac{y_*^2}{R_y^2}+\frac{z_*^2}{R_z^2}\right)\right] \;
\ee
for ${\bf v} = (v,0,0)$. One easily checks that the free 
correlation function, which follows from Eq.~(\ref{Koonin-relat})
or Eq.~(\ref{Koonin-relat-two}), exactly coincides with the formula 
(\ref{corr-free-relat-vx-CM}). To get the correlation function in 
the source rest frame, one performs the Lorentz transformation 
(the correlation function as defined by Eq.~(\ref{def-cov}) is  
a Lorentz scalar) and obtains the formula (\ref{corr-free-relat-vx}). 
Thus, the two ways of `relativization' give the same result for 
non-interacting particles. This is not quite trivial as the time 
dependence of the Koonin formula (\ref{Koonin-relat}) and of the 
explicitly covariant one (\ref{explicit}) look rather different.
Unfortunately, we do not know whether the equivalence of the two
`relativization' schemes holds for interacting particles, as the 
covariant `relativization' is known only for free particles. 

\begin{figure}[t]
\begin{minipage}{5.7cm}
\centering
\includegraphics*[width=5.7cm]{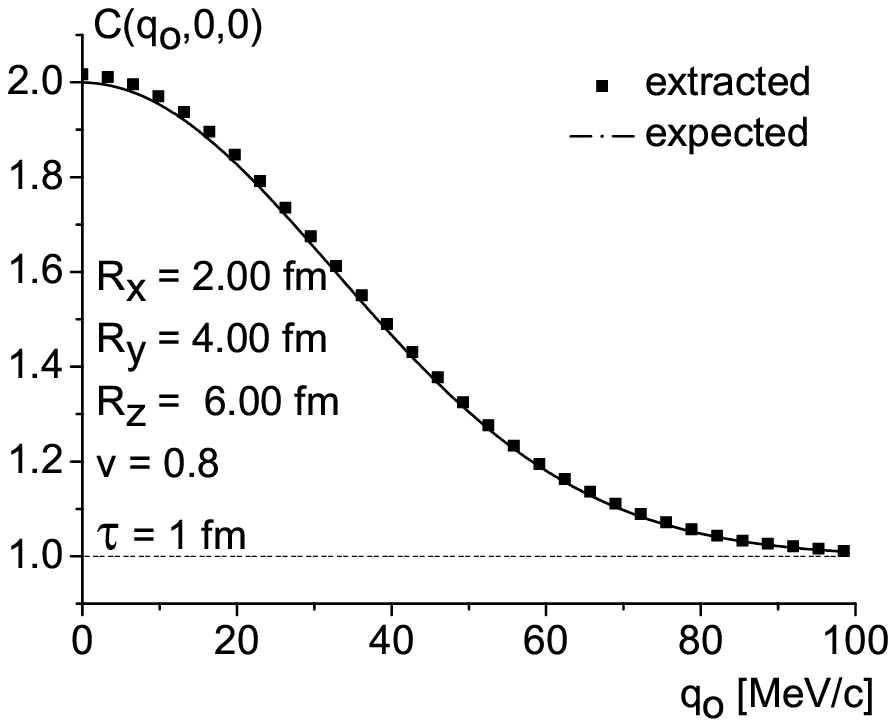}
\end{minipage} \hspace{2mm}
\begin{minipage}{5.7cm}
\centering
\includegraphics*[width=5.7cm]{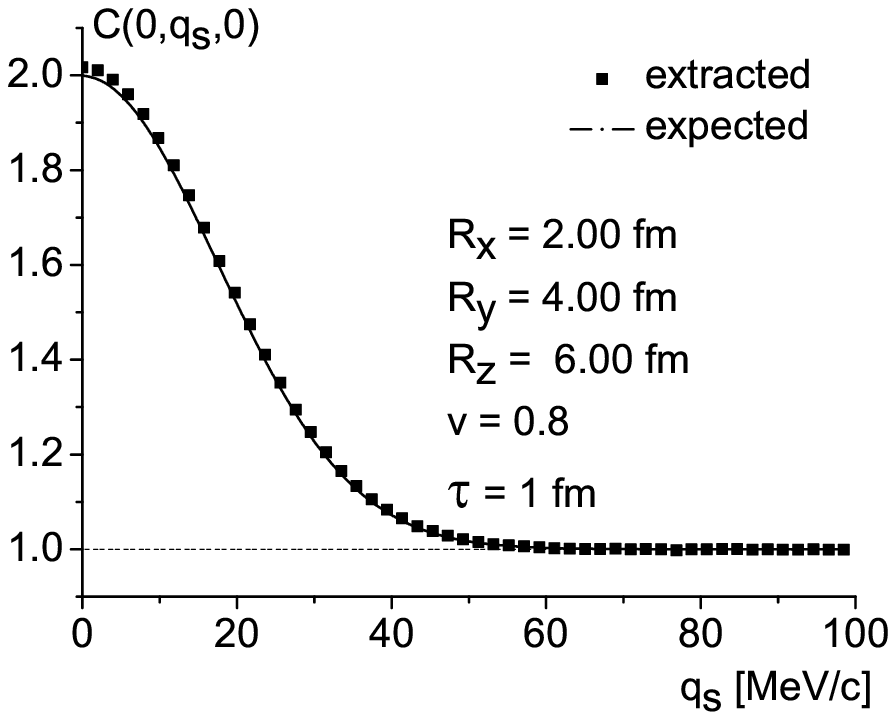}
\end{minipage}\hspace{2mm}
\begin{minipage}{5.7cm}
\centering
\includegraphics*[width=5.7cm]{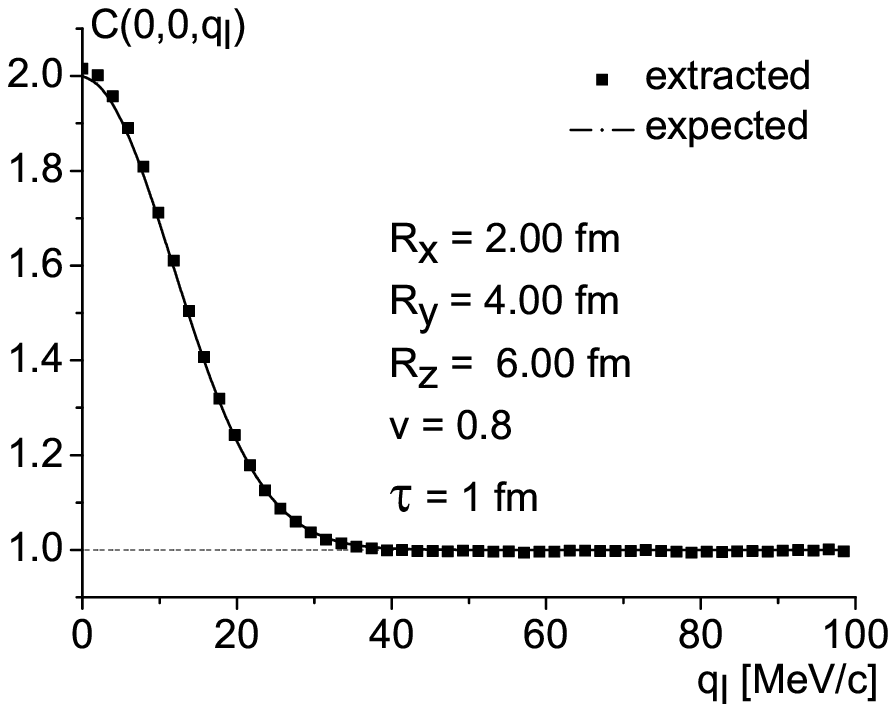}
\end{minipage}
\caption{The $\pi \pi$ Coulomb correlation functions 
$C(q_o,0,0)$ (left panel), $C(0,q_s,0)$ (central panel) 
and  $C(0,0,q_l)$ (right panel) as functions of $q_o$,
$q_s$ or $q_l$, respectively, for the azimuthally 
asymmetric source. The parameters are: $R_x = 2$~fm, 
$R_y = 4$~fm, $R_z = 6$~fm, $\tau=1$~fm 
and the pair velocity is ${\bf v} = (0.8,0,0)$.
The extracted `free' functions are represented by the 
squares while the expected free correlation functions 
correspond to the solid lines.}
\label{fig-aasym-1}
\end{figure}


\section{Coulomb Correlation Functions}
\label{sec-Coulomb}


In this section we compute, using Eq.~(\ref{Koonin-relat-two}), 
the correlation functions of pairs of identical pions or kaons 
interacting due to the Coulomb force. The calculations are performed 
for the anisotropic Gaussian source of finite emission time 
(\ref{source-cov}, \ref{source-matrix}). We use the Bertsch-Pratt 
coordinates \cite{Bertsch:1988db,Pratt:1986cc} {\it out}, {\it side}, 
{\it long}. These are the Cartesian coordinates, where the direction 
{\it long} is chosen along the beam axis ($z$), the {\it out} is 
parallel to the component of the pair momentum ${\bf P}$ which is 
transverse to the beam. The last direction - {\it side} - is along 
the vector product of the {\it out} and {\it long} versors. So, the 
vector ${\bf q}$ is decomposed into the $q_o$, $q_s,$ and $q_l$ 
components. If the particle's velocity is chosen along the axis $x$, 
the out direction coincides with the direction $x$, the side direction 
with $y$ and the long direction with $z$. We note that the correlation 
function of two identical free bosons in the Bertsch-Pratt coordinates 
in the source rest frame is
$$
C({\bf q}) = 1 + 
\exp\big[-4 ( q_o^2R_o^2 + q_s^2R_s^2 + q_l^2R_l^2)\Big],
$$
where $R_o=\sqrt{R_x^2+v^2\tau^2}$,  $R_s = R_y$ and $R_l=R_z$.
As seen, the source lifetime $\tau$ is mixed up with the size parameter
$R_x$. Although experimentalists usually use the parameters $R_o$, $R_s$, 
$R_l$, we use them together with $R_x$, $R_y$, $R_z$ and $\tau$, 
as the lifetime $\tau$ naturally enters theoretical formulas. Since 
the velocity of the pair is chosen along the axis $x$, we always have 
$R_s = R_y$ and $R_l=R_z$.

The effect of Coulomb interaction in femtoscopy can be treated 
analytically or almost analytically under some simplifying 
approximations \cite{Baym:1996wk}. However, we are interested in the 
exact Coulomb correlation functions. Therefore, we use the exact wave 
function. In the case of two non-identical particles interacting due 
to repulsive Coulomb force, the nonrelativistic wave function is well 
known to be \cite{Schiff68}
\be 
\label{coulomb-wave} 
\varphi_{\bf q}({\bf r})
= e^{- {\pi \eta \over 2 q}} \;
\Gamma (1 +i{\eta \over q} ) \; e^{i\bf qr} \;
_1F_1\big(-i{\eta \over q}, 1, i(qr - {\bf qr}) \big) \;,
\ee
where $q \equiv |{\bf q}|$ and $\eta^{-1}$ is the Bohr radius which
for pairs of pions and kaons equals $\eta^{-1}_{\pi} = 388$ fm 
and $\eta^{-1}_{K}=110$ fm, respectively; $_1F_1$ denotes the 
hypergeometric confluent function. When one deals with identical 
bosons, the wave function $\varphi_{\bf q}({\bf r})$ should be 
symmetrized and the modulus of the symmetrized Coulomb wave function 
equals
\ba
\label{modul}
|\varphi_{\bf q}({\bf r})|^2 &=& 
{1 \over 2} \; G(q) \: \Big[
|_1F_1 (-i{\eta \over q}, 1, i(qr - {\bf qr}) )|^2  + 
|_1F_1 (-i{\eta \over q}, 1, i(qr + {\bf qr}) )|^2 
\\[2mm] \nonumber
&+& 2 {\rm Re}\Big(e^{2i{\bf qr}} \: 
_1F_1 (-i{\eta \over q}, 1, i(qr - {\bf qr}) ) \;
_1F_1^*(-i{\eta \over q}, 1, i(qr + {\bf qr}) ) \Big) \Big] \;,
\ea
where $G(q)$ is the so-called Gamov factor defined as
\be 
\label{Gamov}
G(q) = {2 \pi \eta \over q} \,
{1 \over {\rm exp}\big({2 \pi \eta \over q}\big) - 1} \;.
\ee

\begin{figure}[t]
\begin{minipage}{5.7cm}
\centering
\includegraphics*[width=5.7cm]{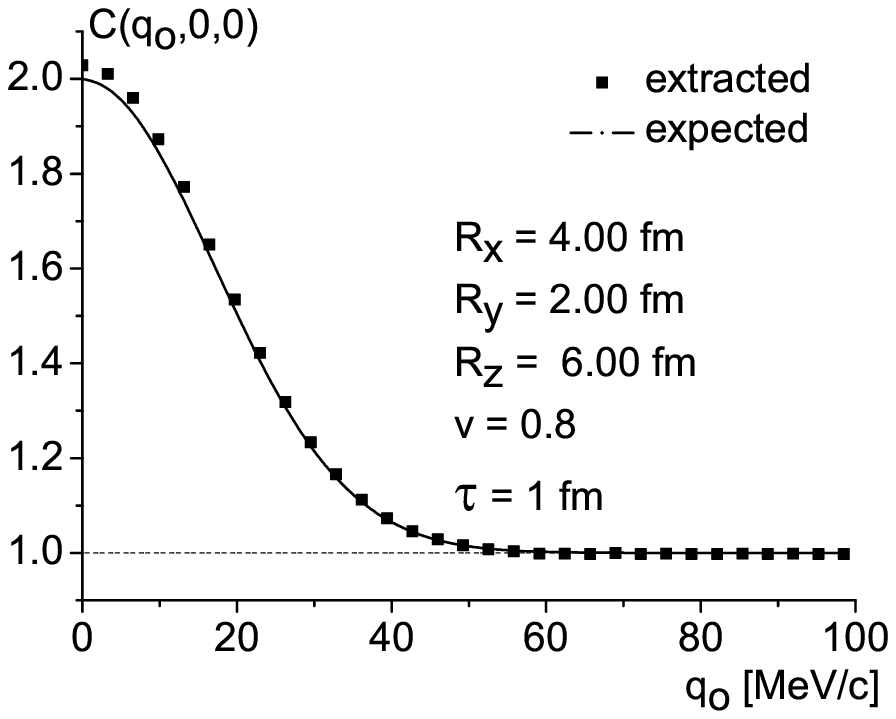}
\end{minipage} \hspace{2mm}
\begin{minipage}{5.7cm}
\centering
\includegraphics*[width=5.7cm]{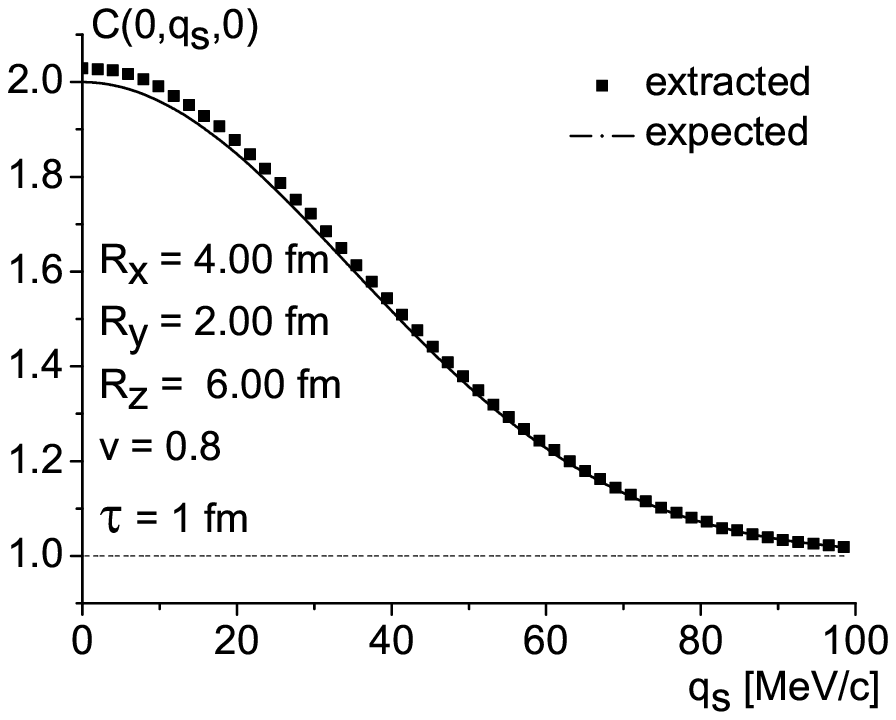}
\end{minipage}\hspace{2mm}
\begin{minipage}{5.7cm}
\centering
\includegraphics*[width=5.7cm]{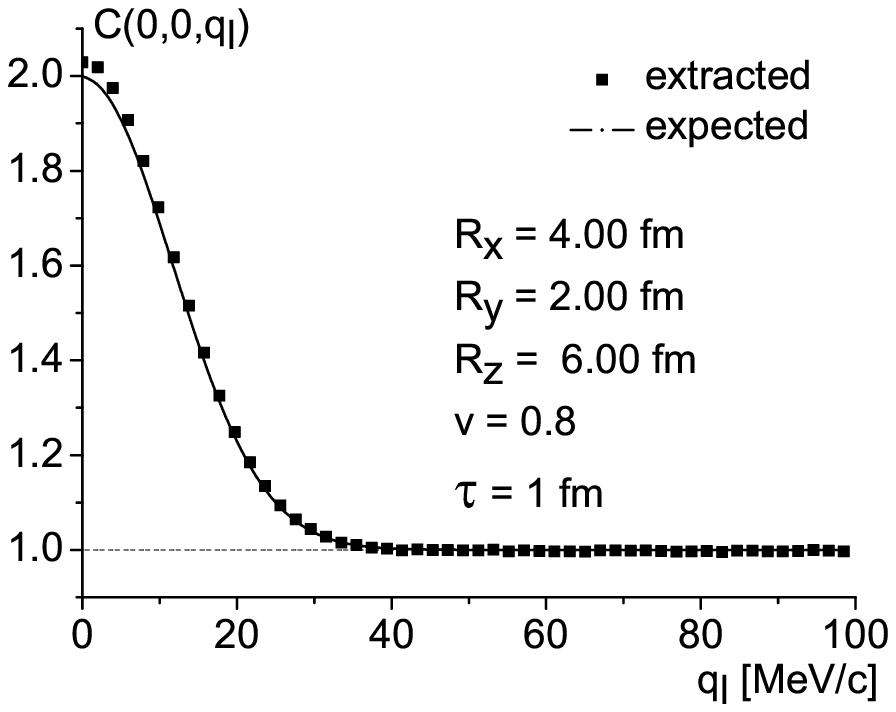}
\end{minipage}
\caption{The $\pi \pi$ Coulomb correlation functions 
$C(q_o,0,0)$ (left panel), $C(0,q_s,0)$ (central panel) 
and  $C(0,0,q_l)$ (right panel) as functions of $q_o$,
$q_s$ or $q_l$, respectively, for the azimuthally 
asymmetric source. The parameters are: $R_x = 4$~fm, 
$R_y = 2$~fm, $R_z = 6$~fm, $\tau=1$~fm and the pair velocity 
is ${\bf v} = (0.8,0,0)$. The extracted `free' functions are 
represented by the squares while the expected free correlation 
functions correspond to the solid lines.}
\label{fig-aasym-2}
\end{figure}

One may wonder whether a multi-particle environment, which
occurs in the final state of relativistic heavy-ion collisions,  
influences the Coulomb potential of the two particles of 
interest. It should be remembered, however, that the particles
are correlated at small relative momenta and thus, they fly 
with similar velocities. Consequently, after the time comparable 
to the source size, the particles with small relative velocity 
appear to be effectively isolated from the rest of many-particle 
system. Therefore, the effect of screening of Coulomb potential 
is expected to be negligible. This qualitative argument is 
confirmed by the calculations presented in \cite{Anchishkin:1996vp}.

Substituting the modulus (\ref{modul}) and the source function
(\ref{source-CM}) into Eq.~(\ref{Koonin-relat-two}), one finds 
the correlation function in the center-of-mass frame which is 
further transformed to the source rest frame. In 
Figs.~\ref{fig-out-side-long-pi-pi} and 
\ref{fig-out-side-long-K-K} we show the correlation functions 
$C(q_o,0,0)$, $C(0,q_s,0)$ and $C(0,0,q_l)$ of identical pions 
and kaons, respectively. The calculations are performed for the 
following values of the source parameters: $R_x = 4$~fm, $R_y = 4$~fm, 
$R_z = 6$~fm, and $\tau= 1,\; 2,\; 3$~fm. The velocity of the 
particle's pair with respect to the source equals $v=0.8$ and 
it is along the axis $x$. As seen, the correlation functions of 
pions and kaons differ sizably due to the different Bohr radii 
of the two systems. The most visible difference appears for the 
function $C(q_o,0,0)$.


\section{The Halo}
\label{sec-halo}


As mentioned in the introduction, the halo \cite{Nickerson:1997js} 
was introduced to explain the fact that, after removing the Coulomb 
effect, the experimentally measured correlation functions are smaller 
than 2 at vanishing relative momentum. The idea of halo assumes that 
only a fraction $f$ ($0\leq f \leq 1$) of particles contributing to 
the correlation function comes from the fireball or core while the 
remaining fraction ($1-f$) originates from long living resonances. 
Then, we have two sources of the particles: the small one - the 
fireball or core - and the big one corresponding to the long living 
resonances. The single-particle source function has two contributions 
\be
D(t,{\bf r}) = f \: D_f(t,{\bf r})+ (1 - f) \: D_h(t,{\bf r}) \;,
\ee
where $D_f(t,{\bf r})$ and $D_h(t,{\bf r})$ represent the fireball 
and halo, respectively. For non-interacting identical bosons, the 
correlation function is
\be
\label{halo-free}
C({\bf q}) = 1 + f^2 e^{-4R_f^2 {\bf q}^2} 
+ (1 - f)^2 e^{-4R_h^2 {\bf q}^2} 
+ 2 f(1-f)e^{-2(R_f^2+R_h^2) {\bf q}^2} \;,
\ee
where both the fireball and halo are assumed to be spherically 
symmetric sources of zero lifetimes; $R_f$ and $R_h$ are the
radii of, respectively, the fireball and the halo. If $R_h$ is 
so large that $R_h^{-1}$ is below an experimental resolution of 
the relative momentum ${\bf q}$, the third and fourth term of 
the correlation function (\ref{halo-free}) are effectively not 
seen, and one claims that $C({\bf q} = 0) = 1 + \lambda$
with $\lambda \equiv f^2 < 1$. 

\begin{figure}[t]
\begin{minipage}{5.7cm}
\centering
\includegraphics*[width=5.7cm]{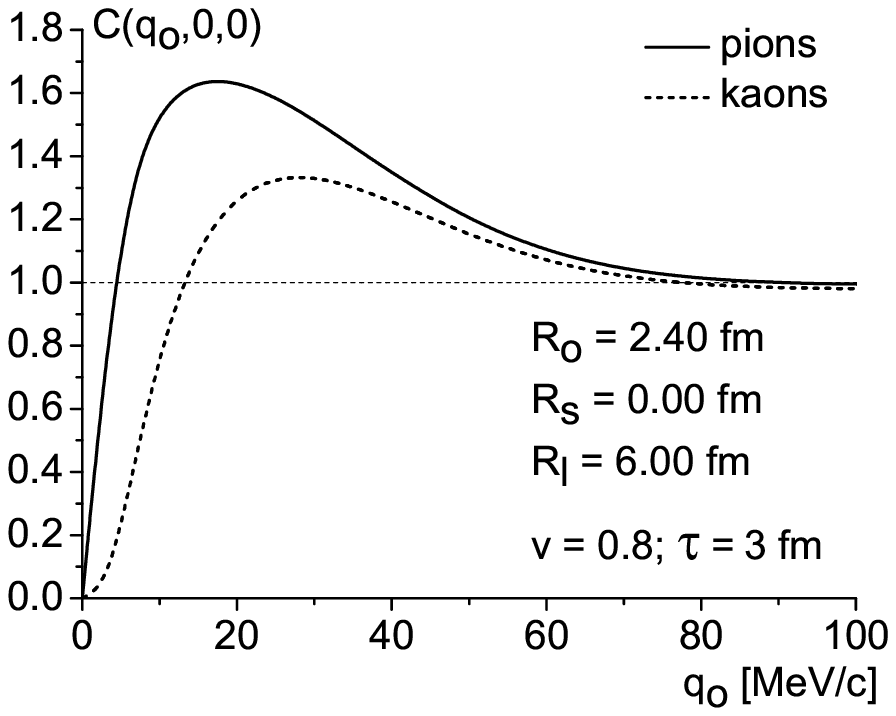}
\end{minipage} \hspace{2mm}
\begin{minipage}{5.7cm}
\centering
\includegraphics*[width=5.7cm]{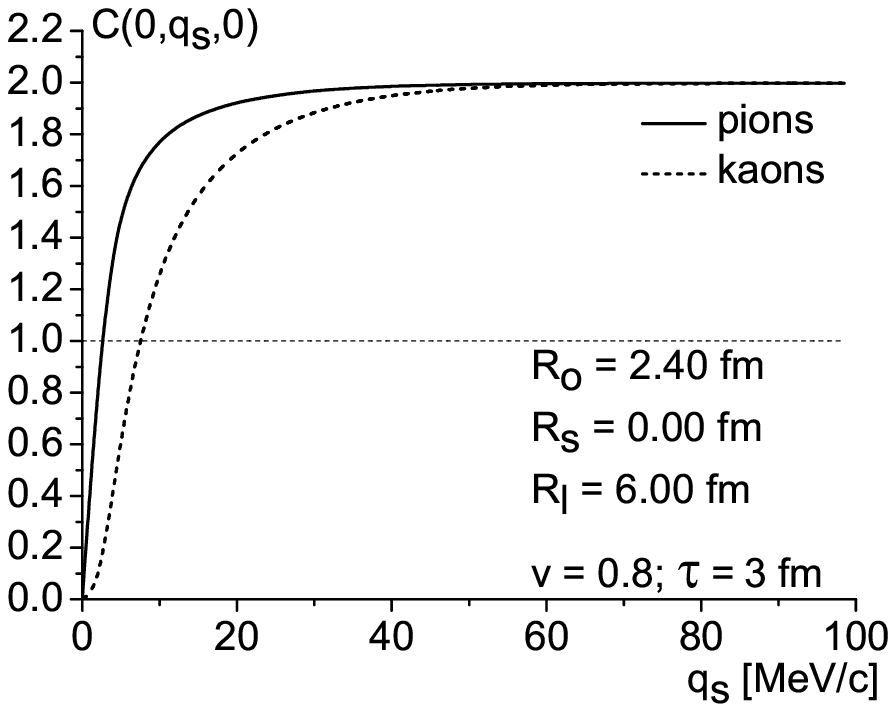}
\end{minipage}\hspace{2mm}
\begin{minipage}{5.7cm}
\centering
\includegraphics*[width=5.7cm]{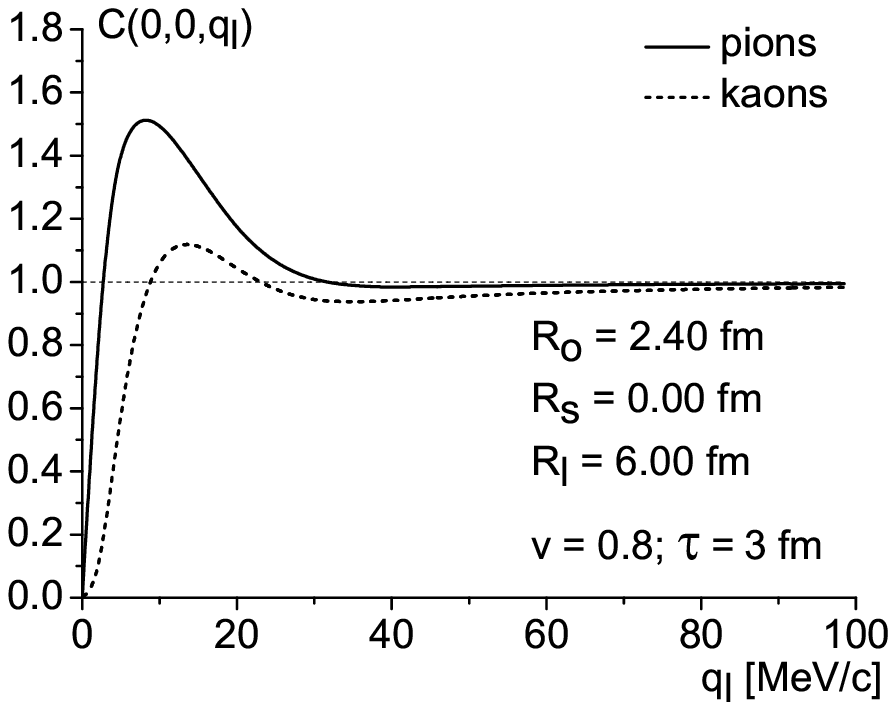}
\end{minipage}
\caption{The $\pi \pi$ and $KK$ Coulomb correlation functions 
$C(q_o,0,0)$ (left panel), $C(0,q_s,0)$ (central panel) 
and  $C(0,0,q_l)$ (right panel) as functions of $q_o$,
$q_s$ or $q_l$, respectively, for the extremely anisotropic 
source. The parameters are $R_x = R_y = 0$, $R_z = 6$~fm, $\tau=3$~fm 
and the pair velocity is ${\bf v} = (0.8,0,0)$.}
\label{fig-coulomb-extreme}
\end{figure}

We have included the halo in our calculations of the $\pi \pi$ 
Coulomb correlation functions. Since the halo represents pions 
from resonances, the source function of the halo, which was  
carefully modeled in \cite{Wiedemann:1996ig}, is approximately
of the exponential form. In our calculations, however, the halo 
source function, as other source functions we use, is of the 
Gaussian form for the reasons explained in the introduction. 
Our simplified treatment of the halo seems to be harmless, as 
the halo influences the correlation function only for $q$ of 
the order $R_h^{-1}$ which are experimentally hardly accessible. 

Our exemplary results are shown in Figs.~\ref{fig-halo-lam} for 
several values of $\lambda$. The fireball is anisotropic with 
$R_x=4$~fm, $R_y=4$~fm, $R_z=6$~fm, $\tau = 3$~fm and $v=0.8$; 
the halo is spherically symmetric, its radius is $R_h =40$~fm 
(as suggested in \cite{Nickerson:1997js}) and its lifetime 
vanishes. 

In principle, a finite spatial extension of the halo implies 
a finite duration of pion emission. However, when the finite 
lifetime of the halo is taken into account, the size of the 
halo in the out direction increases, and the correlation 
function observed in this direction is influenced at even 
smaller momenta than those in the side and long directions. 
In other words, neglecting the finite lifetime of the halo, 
its effect on the correlation function in the out direction
is overestimated not underestimated. We return to this point 
at the end of Sec.~\ref{sec-B-S-with-halo}.


\section{Coulomb Correction without Halo}
\label{sec-B-S-no-halo}


As mentioned in the Introduction, the Coulomb effect is usually 
subtracted from the experimentally measured correlation functions 
by means of the Bowler-Sinyukov procedure. We first note that the 
Coulomb effect is far not small and thus the method to subtract the 
Coulomb effect should be carefully tested. 

In the absence of halo the Bowler-Sinyukov procedure assumes that 
the Coulomb effect can be factorized out, that is the correlation 
function can be expressed as 
\be
\label{corr1}
C({\bf q})=K(q) \: C_{\rm free}({\bf q}) \;,
\ee
where $C_{\rm free}({\bf q})$ is the free correlation function
and $K(q)$ is the correction factor which can be treated 
as the Coulomb correlation function of two nonidentical 
particles of equal masses and charges. The function is, however, 
rather unphysical as the pair velocity vanishes even so the 
calculation is performed in the rest frame of the source where 
the source is assumed to be symmetric and of zero lifetime. 
The correction factor $K(q)$, which is described in detail in 
the Appendix to the paper \cite{Kisiel:2006is}, is computed as
\be
\label{B-S-pop}
K(q) = G(q) \int d^3r \:
D_r({\bf r}) \: |_1F_1(-\frac{i\eta}{q},1,i(qr-{\bf q}{\bf r}))|^2,
\ee
where $G(q)$ is the Gamov factor (\ref{Gamov}) and $D_r({\bf r})$ 
describes the spherically symmetric Gaussian source of zero lifetime 
and of the `effective' radius $R = \sqrt{(R_o^2 + R_s^2 + R_l)/3}$
where $R_o=\sqrt{R_x^2 + v^2 \tau^2}$, $R_s=R_y$ and $R_l=R_z$ are 
the femtoscopic radii obtained from the extracted free correlation 
function. Experimentally $R_o$, $R_s$ and $R_l$ are found fitting 
the measured correlation function $C({\bf q})$ with 
$K(q) \: C_{\rm free}({\bf q})$. In our theoretical analysis, 
$R_x$, $R_y$, $R_z$ and $\tau$ are the actual source parameters 
which enter the source function (\ref{gauss}).

\begin{figure}[t]
\begin{minipage}{5.7cm}
\centering
\includegraphics*[width=5.7cm]{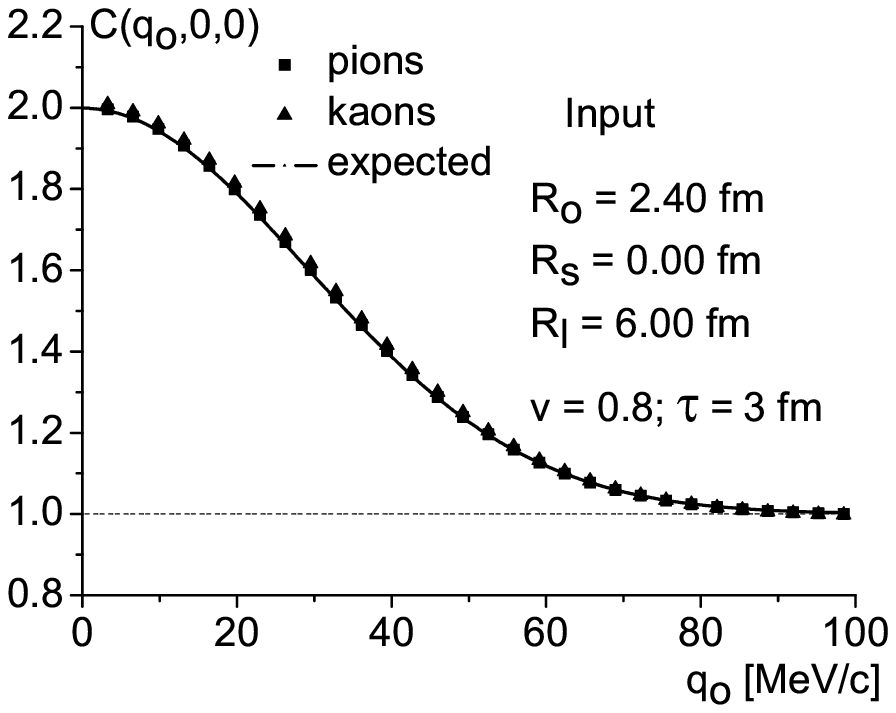}
\end{minipage} \hspace{2mm}
\begin{minipage}{5.7cm}
\centering
\includegraphics*[width=5.7cm]{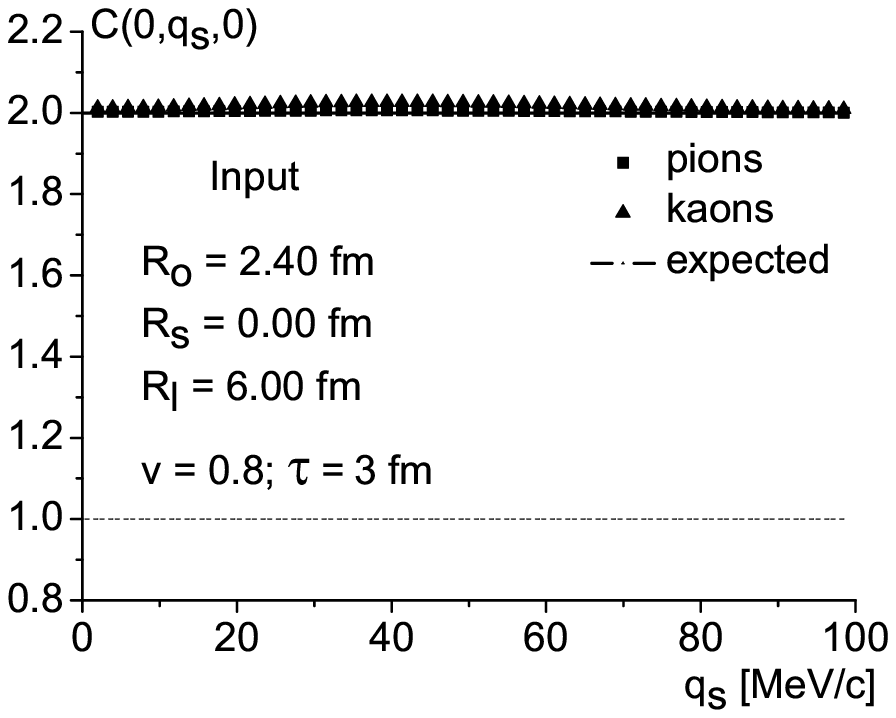}
\end{minipage}\hspace{2mm}
\begin{minipage}{5.7cm}
\centering
\includegraphics*[width=5.7cm]{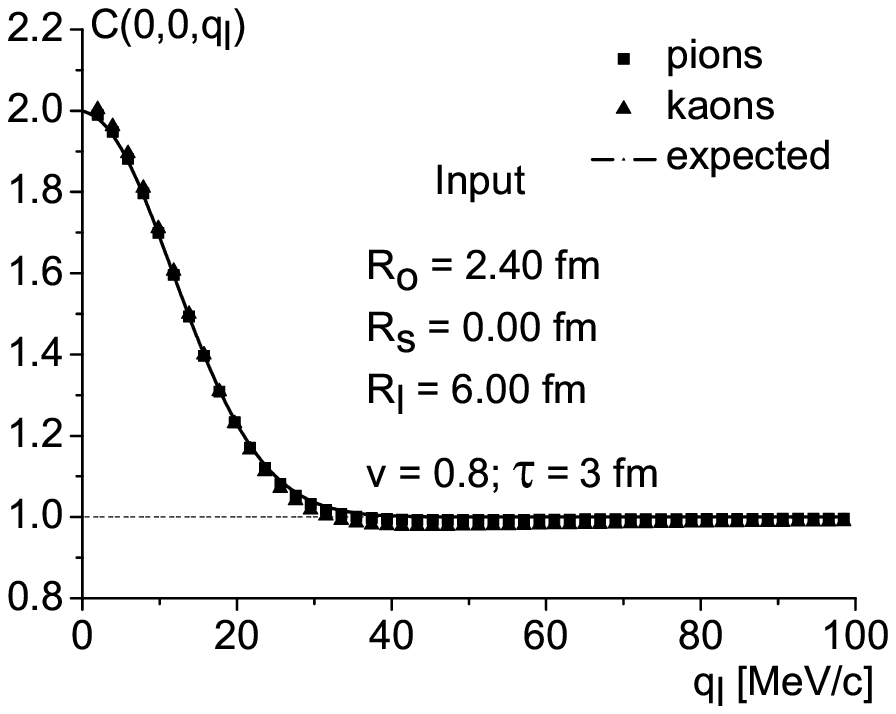}
\end{minipage}
\caption{The $\pi \pi$ and $KK$ `free' correlation 
functions $C(q_o,0,0)$ (left panel), $C(0,q_s,0)$ 
(central panel) and  $C(0,0,q_l)$ (right panel) as 
functions of $q_o$, $q_s$ or $q_l$, respectively, for 
the extremely anisotropic source. The parameters are: 
$R_x = R_y = 0$, $R_z = 6$~fm, $\tau=3$~fm and the pair 
velocity is ${\bf v} = (0.8,0,0)$.The extracted `free' 
functions are represented by the squares for pions
and by triangles for kaons; the expected free correlation functions 
correspond to the solid lines.}
\label{fig-free-extreme}
\end{figure}

Using the parabolic coordinates $\xi_+ \equiv r+z$,
$\xi_- \equiv r-z$ and the azimuthal angle $\phi$, the relative source
function of isotropic Gaussian source of zero lifetime is 
\be
\label{para-source}
D_r({\xi_+,\xi_-,\phi}) = \frac{1}{8 \pi^{3/2}R^3}
\exp \Big(-\frac{(\xi_+ + \xi_-)^2}{16R^2}\Big) \;,
\ee
which substituted into Eq.~(\ref{B-S-pop}) gives
\be
\label{K-almost-excat}
K(q) =  \frac{G(q)}{16\pi^{1/2}R^3}
\int_0^{\infty} d\xi_+ \int_0^{\infty}d\xi_- (\xi_+ + \xi_-)
\Big|{_1}F_1\Big(-\frac{i\eta}{q},1,iq\xi_-\Big)\Big|^2
\exp \Big(-\frac{(\xi_+ + \xi_-)^2}{16R^2} \Big) \;.
\ee
The trivial integral over $\phi$ has been performed in 
Eq.~(\ref{K-almost-excat}). Since the confluent hypergeometric 
function does not depend on $\xi_+$, the integral over $\xi_+$ 
can be easily performed and one obtains
\be
\label{K-exact}
K(q) =  \frac{G(q)}{2\pi^{1/2}R}
\int_0^{\infty} d\xi_- 
\Big|{_1}F_1\Big(-\frac{i\eta}{q},1,iq\xi_-\Big)\Big|^2
\exp \Big(-\frac{\xi_-^2}{16R^2} \Big) \;,
\ee
where the integral over $\xi_-$ is usually computed numerically. 
However, observing that the source size is always much smaller 
than the Bohr radius of the particles of interest, one derives 
the approximate expression of the hyperbolic confluent function 
(\ref{approx-2}) which is discussed in Appendix B. With the 
formula (\ref{approx-3}) the integration can be performed 
analytically and the correction factor equals
\be
\label{B-S-pop-appr}
K(q) = G(q)\left[1 + \frac{8\eta R}{\sqrt{\pi}} \:
{_{2}F_2} \left(\frac{1}{2},1;\frac{3}{2},\frac{3}{2};
-4q^2R^2\right)\right] \,.
\ee

In Fig.~\ref{popandapro} we show the correction factor $K(q)$
for pions and kaons computed from the exact formula (\ref{B-S-pop}) 
and the approximate one (\ref{B-S-pop-appr}). To make the difference
more visible (note the vertical scale) the correction factor is 
divided by the Gamov factor which strongly varies with $q$. 
One sees that the approximation (\ref{B-S-pop-appr}) is very 
accurate for pions and it is less accurate for kaons. For this 
reason the expression (\ref{B-S-pop-appr}) is used only for pions. 
Fig.~\ref{popandapro} also shows that the correction factor is 
heavily dominated by the Gamov factor that is $K(q)/G(q)$ differs 
very little from unity.

Once we are able to compute the exact Coulomb correlation functions 
for an anisotropic source of finite lifetime, we can test 
whether the free correlation function obtained by means of the 
Bowler-Sinyukov equation (\ref{corr1}) properly reproduces the 
actual free correlation function. 

The free correlation functions, which are obtained using 
Eq.~ (\ref{corr1}), are shown in Fig.~\ref{fig-free} for pions 
in Fig.~\ref{fig-free-KK} for kaons. The extracted functions 
are compared to the expected correlation functions of noninteracting 
bosons for the given source. As seen, the free correlation 
function is almost exactly reproduced in the case of pions 
while in the case of kaons the reproduction is less accurate.
Similar results are found as long as the source radii are
much smaller than the Bohr radius of particles of interest.


\section{Azimuthally sensitive femtoscopy}


In the previous sections we discussed the particle sources of cylindrical
(azimuthal) symmetry ($R_x = R_y$). The sources created in non-central 
collisions are not azimuthally symmetric but the symmetry is usually 
restored due to the averaging over impact parameter orientation.  
The cylindrically asymmetric sources are observable, if the reaction 
plane is determined. The azimuthally sensitive femtoscopy was developed 
\cite{Adams:2003ra,Lisa:2003ze} and when applied to experimental data 
it showed an expected dependence of the source radii on the emission 
angle with respect to the reaction plane. Since the Coulomb effects
were removed from the data by means of the Bowler-Sinyukov procedure
in the experimental studies \cite{Adams:2003ra,Lisa:2003ze},
we test in this section the procedure for the case azimuthally 
asymmetric sources. 

\begin{figure}[t]
\begin{minipage}{5.7cm}
\centering
\includegraphics*[width=5.7cm]{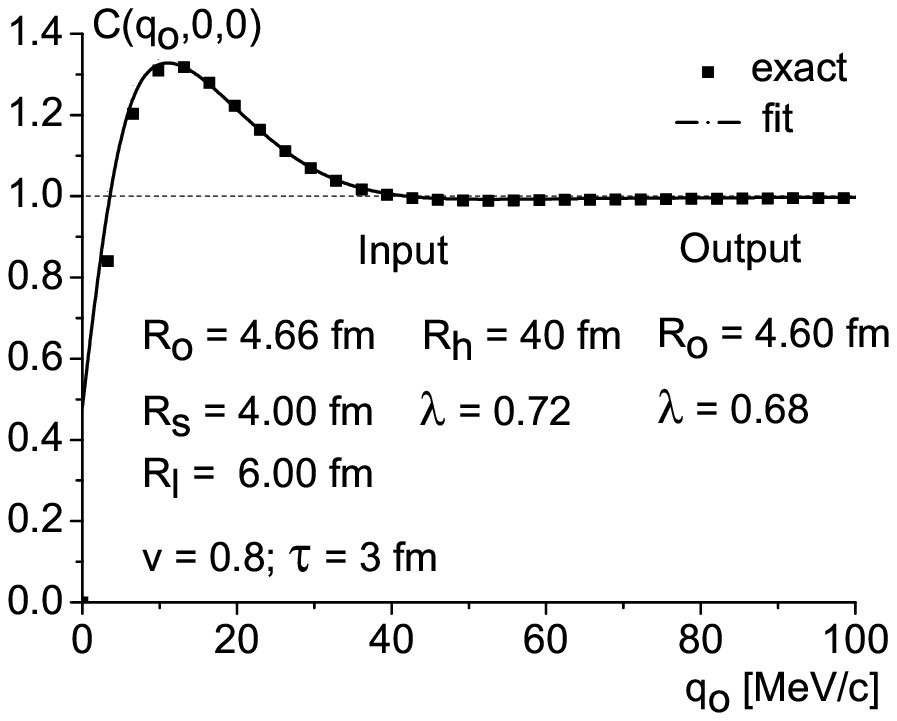}
\end{minipage} \hspace{2mm}
\begin{minipage}{5.7cm}
\centering
\includegraphics*[width=5.7cm]{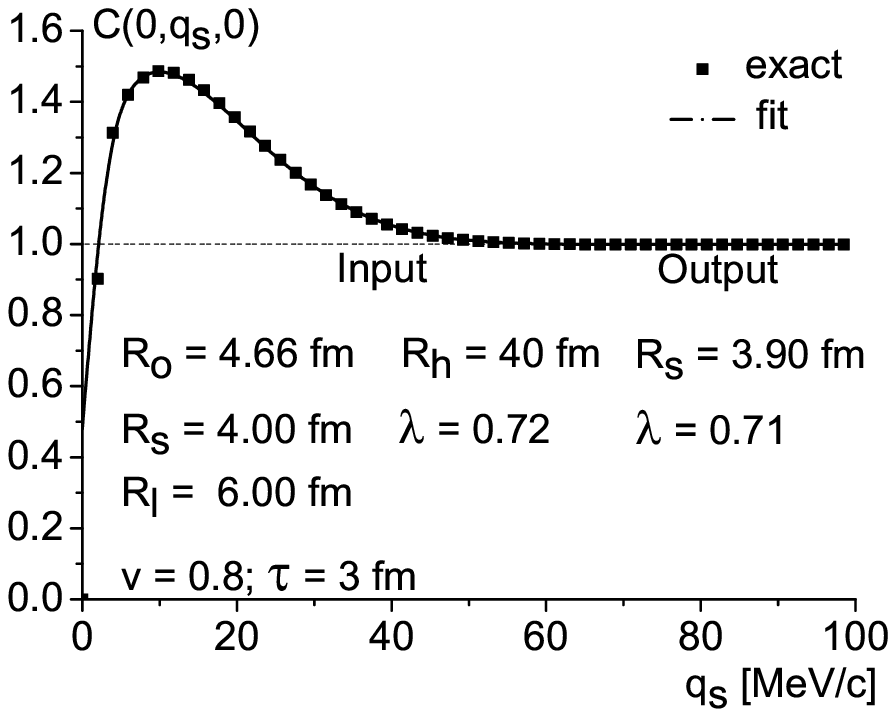}
\end{minipage}\hspace{2mm}
\begin{minipage}{5.7cm}
\centering
\includegraphics*[width=5.7cm]{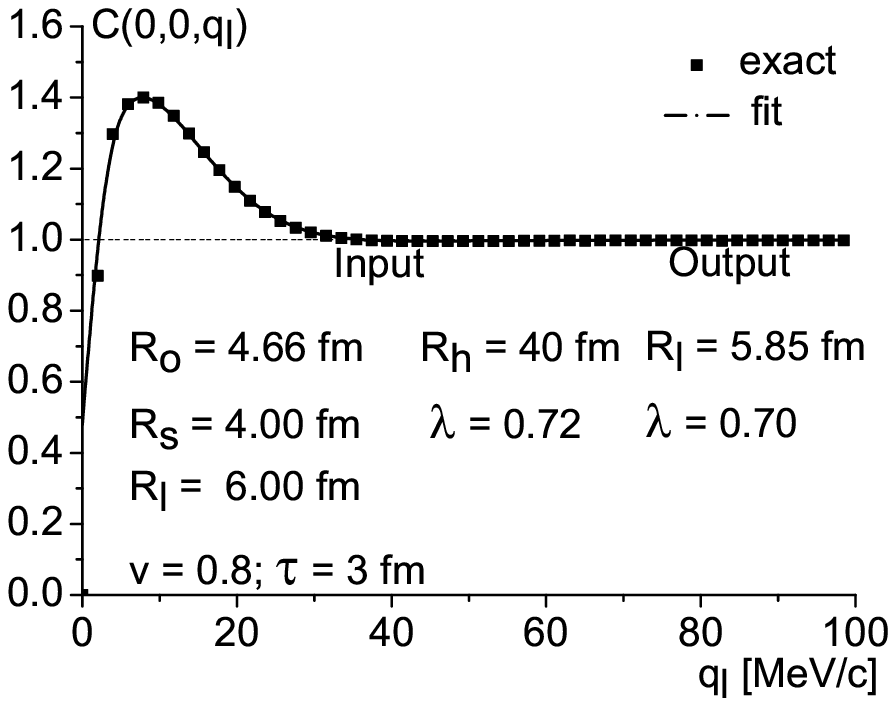}
\end{minipage}
\caption{The $\pi \pi$ Coulomb correlation 
functions $C(q_o,0,0)$ (left panel), $C(0,q_s,0)$ (central panel) 
and  $C(0,0,q_l)$ (right panel) as functions of $q_o$,
$q_s$ or $q_l$, respectively. The fireball and halo parameters 
are $R_x = 4$~fm, $R_y = 4$~fm, $R_z = 6$~fm, $\tau=3$~fm,
$R_h = 40$~fm and the pair velocity is ${\bf v} = (0.8,0,0)$. 
The exact Coulomb correlation functions are shown with the 
squares and the Coulomb correlation functions fitted with 
the dilution formula (\ref{dilution}) are represented by the 
solid lines.}
\label{fig-dilution-fit}
\end{figure}

We do not study a full dependence of the correlation function
on the azimuthal emission angle but we consider two extreme cases.
We stick to our convention that particles are always emitted 
along the axis $x$ but $R_x \not= R_y$. In Fig.~\ref{fig-aasym-1}
we show the pion correlation functions $C(q_o,0,0)$, $C(0,q_s,0)$  
and  $C(0,0,q_l)$ for $R_x = 2$~fm, $R_y = 4$~fm and in 
Fig.~\ref{fig-aasym-2} the case $R_x = 4$~fm, $R_y = 2$~fm
is illustrated. The remaining parameters are  $R_z = 6$~fm, 
$\tau = 1$~fm, ${\bf v} = (0.8,0,0)$. As seen, the Bowler-Sinyukov 
procedure works very well in both cases. 


\section{Extremely Anisotropic Source}


To establish limitations of the Bowler-Sinyukov procedure in
the absence of halo, we have considered an extremely anisotropic
source where $R_z$ is much larger than $R_x$ and $R_y$ as well
as $R_z \gg \tau$. The source function is found from 
Eq.~(\ref{source-CM}) by taking the limits $R_y \rightarrow 0$, 
$R_x \rightarrow 0$ and $\tau \rightarrow 0$. Thus, one finds
\be
\label{source-CM-delta2}
D_r({\bf r}_*)= \frac{1}{2\pi^{1/2}R_z}\exp\left[-\frac{z_*^2}{4R_z^2}\right]\delta(x_*)\delta(y_*) \;.
\ee   
A paradoxical feature of this source function is that the
information about the velocity $v$ of the pair's center-of-mass frame 
with respect to the source has disappeared. Thus, we have the same
source function in the source rest frame and in the center-of-mass 
frame of the pair. However, when we transform the correlation 
function from the pair center-of-mass to the source rest frame the 
pair velocity enters. The advantage of Eq.~(\ref{source-CM-delta2}) 
is that the calculations can be performed almost analytically. 

Substituting the source function (\ref{source-CM-delta2}) 
into Eq.~(\ref{Koonin-relat-two}), one finds 
\ba 
\label{coulomb-out-extreme}
C(q_x^*,0,0) &=& \frac{2G(q_x^*)}{\sqrt{\pi}R_z}
\int_0^{\infty}dz
|_1F_1(-i\frac{\eta}{q_x^*},1,iq_x^* z)|^2 
e^{-\frac{z^2}{4R_z^2}} \;,
\\[2mm]
\label{coulomb-side-extreme}
C(0,q_y^*,0) &=& \frac{2G(q_y^*)}{\sqrt{\pi}R_z}
\int_0^{\infty}dz
|_1F_1(-i\frac{\eta}{q_y^*},1,iq_y^* z)|^2 
e^{-\frac{z^2}{4R_z^2}} \;,
\\[2mm]
\label{coulomb-long-extreme}
C(0,0,q_z^*)&=& \frac{G(q_z^*)}{2\sqrt{\pi}R_z}\int_0^{\infty}dz
|_1F_1(-i\frac{\eta}{q_z^*},1,2iq_z^*z)|^2 
e^{-\frac{z^2}{4R_z^2}} 
\\[2mm]\nonumber
&&+ 2\int_0^{\infty}dz
\left[{\rm Re}\left(e^{-2iq_z^*z}
 {_1F_1}(-i\frac{\eta}{q_z^*},1,2iq_z^*z)\right)+1\right]
e^{-\frac{z^2}{4R_z^2}} \;.
\ea   
For pions the integration over $z$ is performed using the 
approximate expression of the hypergeometric confluent function 
(\ref{approx-2}). Thus, we find
\ban 
C(q_x^*,0,0) &=& 
2G(q_x^*) \left[1+\frac{4\eta R_z}{\sqrt{\pi}}
{_2F_2}(\frac{1}{2},1;\frac{3}{2},\frac{3}{2};-4q_x^{*2} R_z^2)
\right] \;,
\\[2mm]
C(0,q_y^*,0) &=& 
2G(q_y^*) \left[1+\frac{4\eta R_z}{\sqrt{\pi}}
{_2F_2}(\frac{1}{2},1;\frac{3}{2},\frac{3}{2};-4q_y^{*2} R_z^2)
\right]
\;,
\\[2mm]
C(0,0,q_z^*)&=&  
G(q_z^*)\left(1+e^{-4q_z^{*2}R_z^2}\right)
+G(q_z^*)\frac{4\eta R_z}{\sqrt{\pi}}
{_2F_2}(\frac{1}{2},1;\frac{3}{2},\frac{3}{2};-4q_z^{*2}R_z^2)
\\[2mm]\nonumber
&\times& 
\left(1+\sum_0^{\infty}(-1)^n 
\frac{(4q_z^*R_z)^{2n}n!}{(2n)!}
\frac{{_2F_2}(\frac{1}{2},n+1;\frac{3}{2},\frac{3}{2};-4q_z^{*2}R_z^2)}
{{_2F_2}(\frac{1}{2},1;\frac{3}{2},\frac{3}{2};-4q_z^{*2}R_z^2)}\right)
\;.
\ean   

The Bowler-Sinyukov procedure works very well for the correlation
function computed with the extremely anisotropic source function 
(\ref{source-CM-delta2}) with $R_z=6$~fm. To make the test of the
procedure even more challenging, we have considered an extremely
anisotropic source of finite lifetime. The Coulomb correlation 
functions of pions and kaons, which are computed for the source of 
$R_x = R_y = 0$, $R_z = 6$~fm, $\tau=3$~fm and $v=0.8$, are shown 
in Fig.~\ref{fig-coulomb-extreme}. The `free' correlation 
functions, which are presented in Fig.~\ref{fig-free-extreme}, are 
obtained from the correlation functions shown in 
Fig.~\ref{fig-coulomb-extreme} by dividing them by the correction 
factor $K(q)$. The factor is computed for the averaged radius 
$R = \sqrt{(R_0^2 +R_l^2)/3}$. As seen, the Bowler-Sinyukov procedure works 
very well for both pions and kaons. In particular, 
$C_{\rm free}(0,q_s,0) = C(0,q_s,0)/K(q) \approx 2$ as expected.


\section{Coulomb Correction with Halo}
\label{sec-B-S-with-halo}


\begin{figure}[t]
\begin{minipage}{5.7cm}
\centering
\includegraphics*[width=5.7cm]{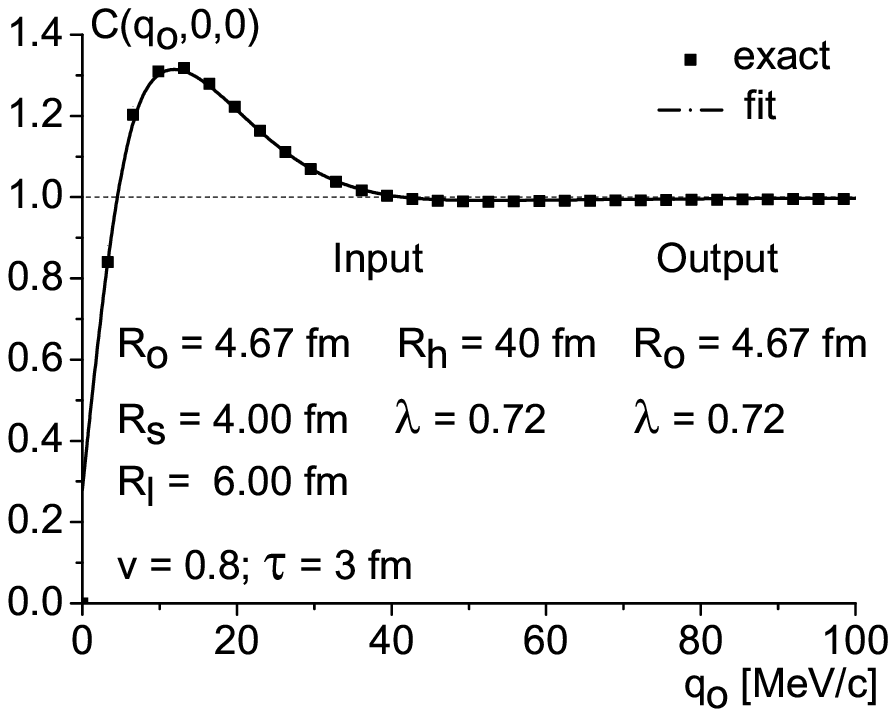}
\end{minipage} \hspace{2mm}
\begin{minipage}{5.7cm}
\centering
\includegraphics*[width=5.7cm]{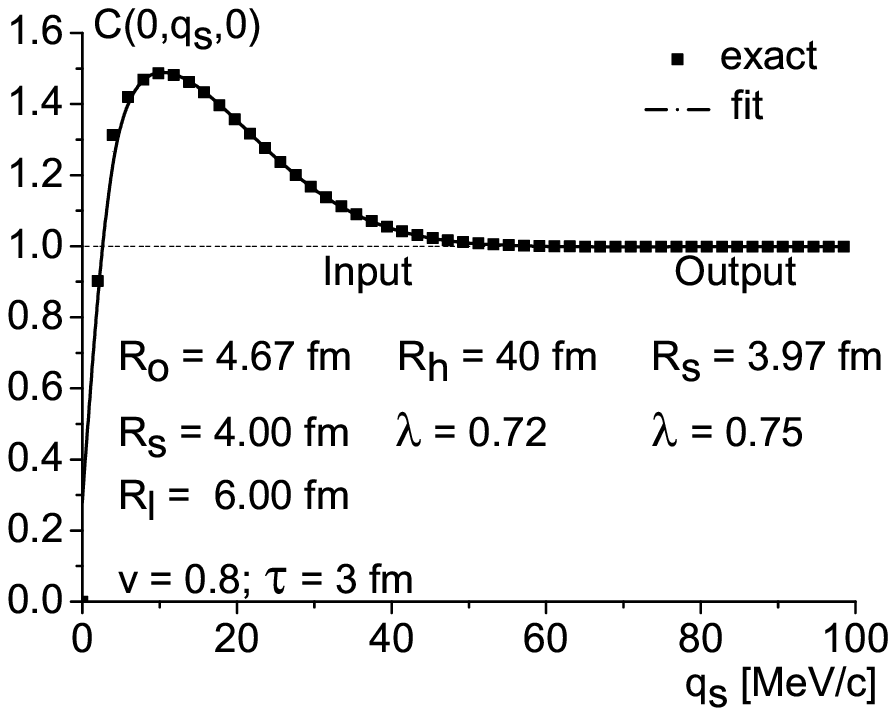}
\end{minipage}\hspace{2mm}
\begin{minipage}{5.7cm}
\centering
\includegraphics*[width=5.7cm]{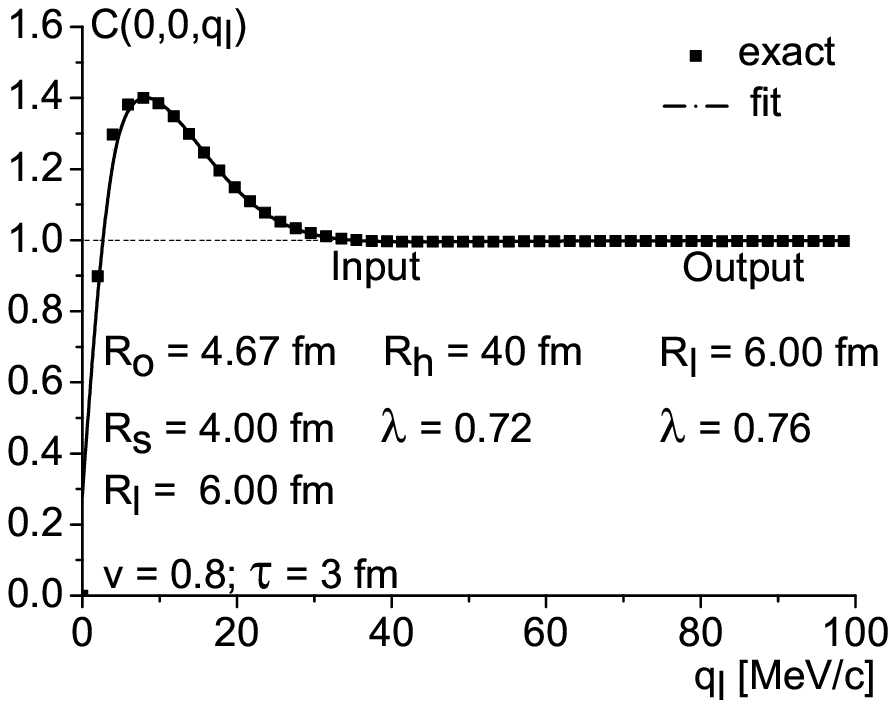}
\end{minipage}
\caption{The $\pi \pi$ Coulomb correlation 
functions $C(q_o,0,0)$ (left panel), $C(0,q_s,0)$ (central panel) 
and  $C(0,0,q_l)$ (right panel) as functions of $q_o$,
$q_s$ or $q_l$, respectively. The fireball and halo parameters 
are $R_x = 4$~fm, $R_y = 4$~fm, $R_z = 6$~fm, $\tau=3$~fm,
$R_h = 40$~fm and the pair velocity is ${\bf v} = (0.8,0,0)$. 
The exact Coulomb correlation functions are shown with the 
squares and the Coulomb correlation functions fitted with 
the Bowler-Sinykov formula (\ref{B-S}) are represented by 
the solid lines.}
\label{fig-B-S-fit}
\end{figure}

The procedure to eliminate the Coulomb interaction is more complex 
when the halo is taken into account. We test two versions of the 
procedure which, following the STAR Collaboration \cite{Adams:2004yc},
we call the `dilution' method and the `proper Bowler-Sinyukov' one. The experimentally measured correlation functions $C({\bf q})$ are fitted as
\be
\label{dilution}
C({\bf q}) = \big(1-\lambda +\lambda K(q) \big)
\Big[1+ \lambda \big( C_{\rm free}({\bf q})-1\big) \Big] \;,
\ee
in the case of the dilution method and
\be
\label{B-S}
C({\bf q}) = 1-\lambda +\lambda K(q)C_{\rm free}({\bf q}) \;,
\ee
in the case of the Bolwer-Sinykov method.

\begin{figure}[t]
\begin{minipage}{5.7cm}
\centering
\includegraphics*[width=5.7cm]{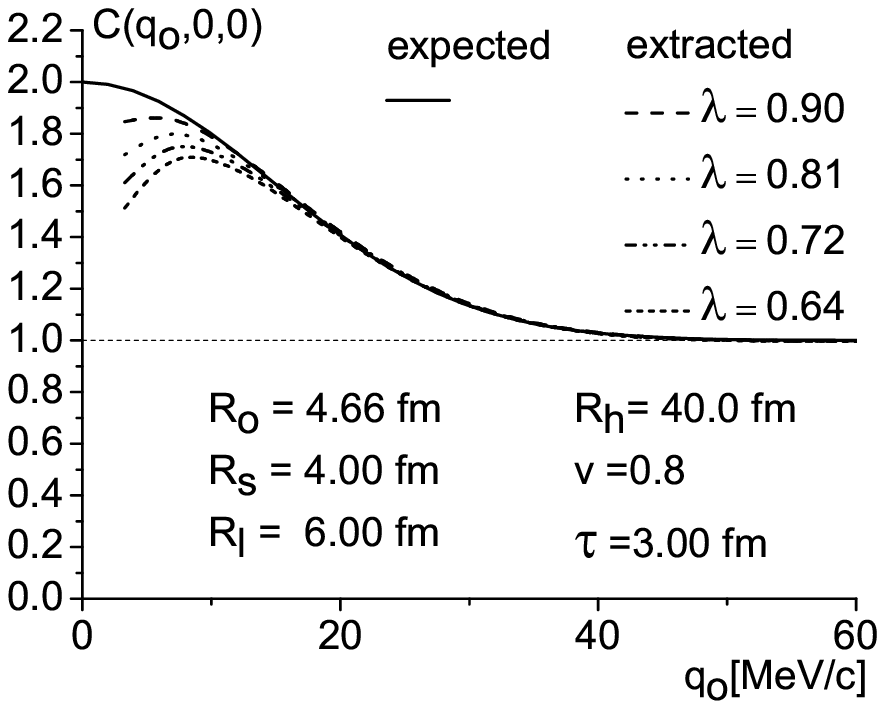}
\end{minipage} \hspace{2mm}
\begin{minipage}{5.7cm}
\centering
\includegraphics*[width=5.7cm]{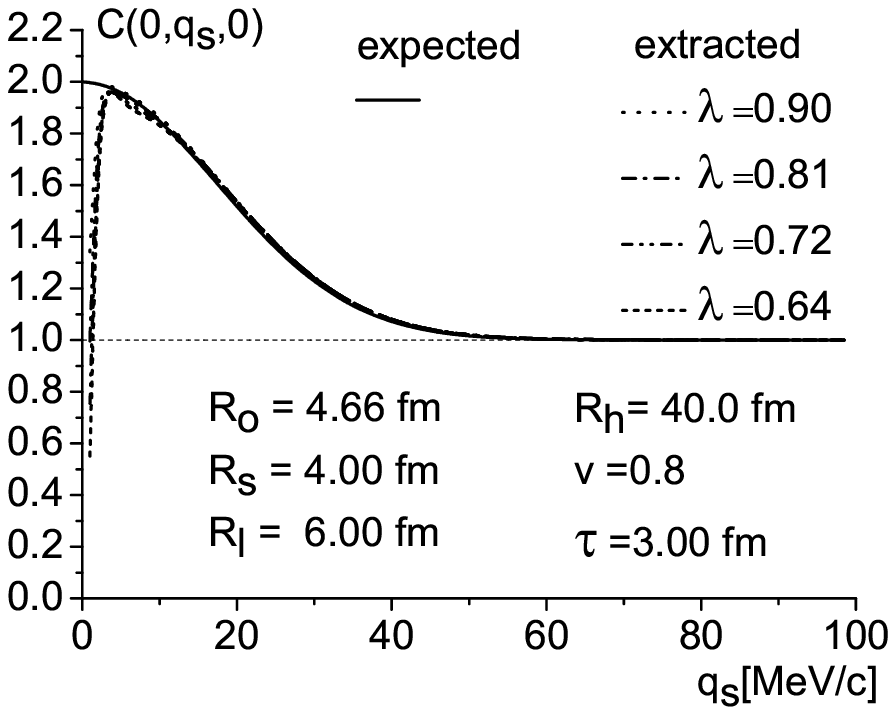}
\end{minipage}\hspace{2mm}
\begin{minipage}{5.7cm}
\centering
\includegraphics*[width=5.7cm]{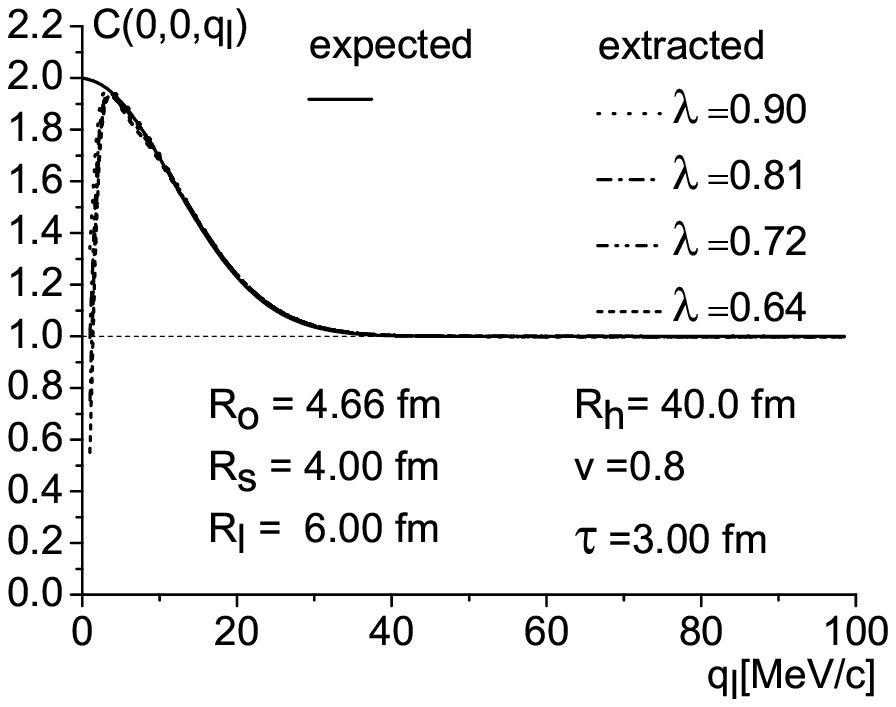}
\end{minipage}
\caption{The $\pi \pi$ free correlation functions 
$C(q_o,0,0)$ (left panel), $C(0,q_s,0)$ (central panel) 
and  $C(0,0,q_l)$ (right panel) as functions of $q_o$,
$q_s$ or $q_l$, respectively. The fireball and halo 
parameters are $R_x = 4$~fm, $R_y = 4$~fm, $R_z = 6$~fm, 
$\tau=3$~fm, $R_h = 40$~fm and the pair velocity is 
${\bf v} = (0.8,0,0)$. The `free' correlation functions are
extracted by means of the dilution method for several values 
of $\lambda$. The actual free correlation functions are
shown by the solid lines.}
\label{fig-dilution-free}
\end{figure}

The Coulomb correlation functions fitted according to the dilution
(\ref{dilution}) and Bowler-Sinyukov (\ref{B-S}) formulas are shown 
in Fig.~\ref{fig-dilution-fit}, \ref{fig-B-S-fit}, respectively. 
The source parameters are given in the figures. The {\em input}
parameters are those used in the computation of Coulomb correlation
functions: the fireball parameters are $R_x = 4$~fm, $R_y = 4$~fm, 
$R_z = 6$~fm and $\tau=3$~fm; the halo is of zero lifetime of the
radius $R_h = 40$~fm, and the pair velocity is ${\bf v} = (0.8,0,0)$.
The {\em output} parameters are obtained from the fit. As seen,
they only slightly deviate from the input parameters. Since the formulas 
(\ref{dilution}, \ref{B-S}) do not work at $q \le 1/R_h$, we perform 
the fit in the domain of $q_o$, $q_s$ or $q_l$, respectively, bigger 
than 6 MeV. As seen the Coulomb correlation functions are fitted very 
accurately with both the dilution (\ref{dilution}) and Bowler-Sinyukov 
(\ref{B-S}) formulas.

\begin{figure}[t]
\begin{minipage}{5.7cm}
\centering
\includegraphics*[width=5.7cm]{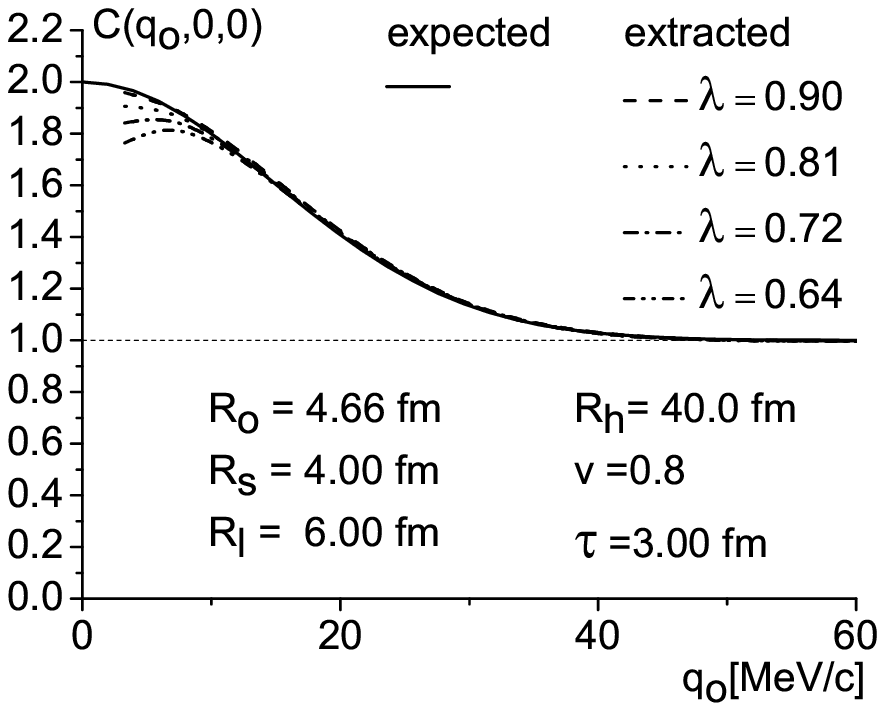}
\end{minipage} \hspace{2mm}
\begin{minipage}{5.7cm}
\centering
\includegraphics*[width=5.7cm]{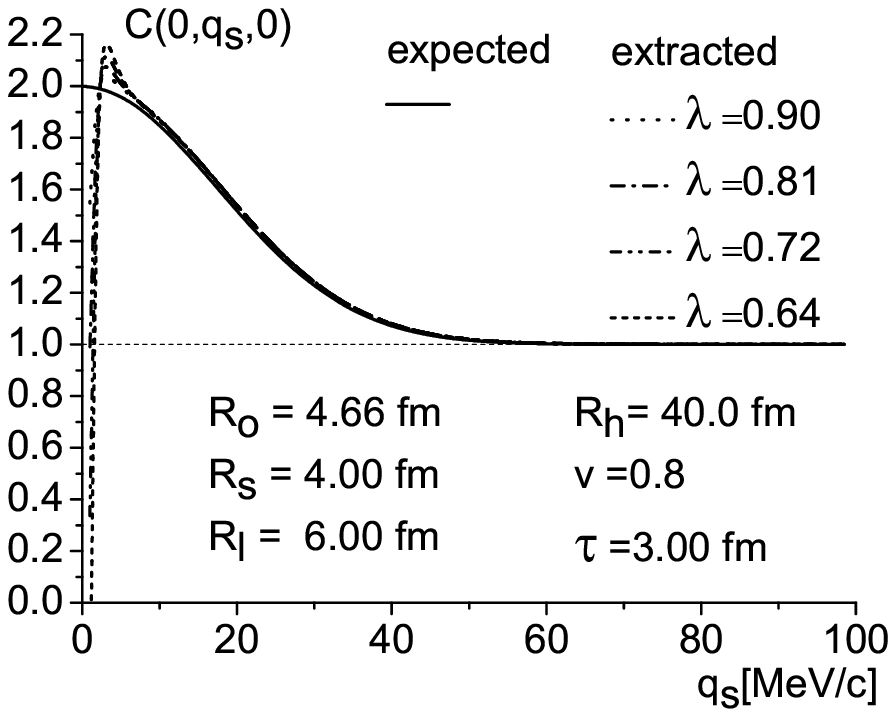}
\end{minipage}\hspace{2mm}
\begin{minipage}{5.7cm}
\centering
\includegraphics*[width=5.7cm]{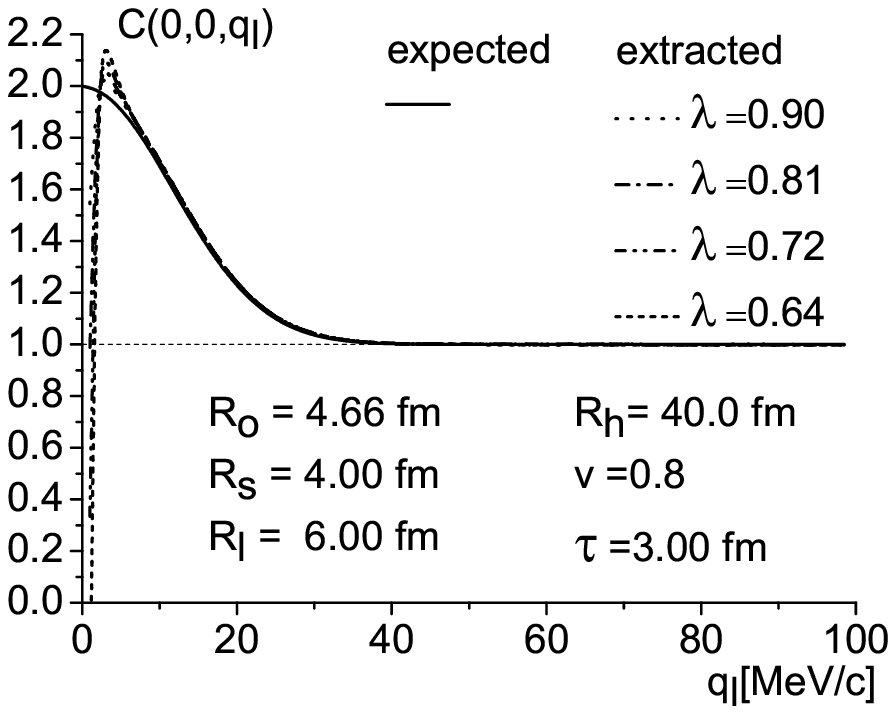}
\end{minipage}
\caption{The $\pi \pi$ `free' correlation functions 
$C(q_o,0,0)$ (left panel), $C(0,q_s,0)$ (central panel) 
and  $C(0,0,q_l)$ (right panel) as functions of $q_o$,
$q_s$ or $q_l$, respectively. The fireball and halo parameters 
are $R_x = 4$~fm, $R_y = 4$~fm, $R_z = 6$~fm, $\tau=3$~fm,
$R_h = 40$~fm and the pair velocity is ${\bf v} = (0.8,0,0)$. 
The `free' correlation functions are extracted by means of the 
Bowler-Sinykov method for several values of $\lambda$. The 
actual free functions are shown by the solid lines.}
\label{fig-B-S-free}
\end{figure}

The `free' correlation functions extracted according to the 
dilution (\ref{dilution}) and Bowler-Sinyukov (\ref{B-S}) 
formulas are shown in Fig.~\ref{fig-dilution-free}, \ref{fig-B-S-free}, respectively. The expected free functions are also shown for 
comparison. It is important to note that the parameter $\lambda$ is 
assumed here to be known that is the actual value of $\lambda$ enters 
the formula (\ref{dilution}) or (\ref{B-S}). As seen, the extracted 
correlation functions are distorted at small relative momenta and 
the distortions grow with $\lambda$. However, the widths of the 
correlation functions are unaltered and so are the source parameters. 

The experimentally obtained `free' functions, which are shown 
e.g. in Fig.~4 from \cite{Adams:2004yc}, do not reveal the dip 
at small ${\bf q}$ seen Fig.~13 and 14. We note, however, that 
the experimental correlations function in, say, out direction 
are not of the form $C(q_o,0,0)$ but rather 
$\int dq_s dq_l C(q_o,q_s,q_l)$ and the domain of truly small 
${\bf q}$ is not seen. 

As mentioned at the end of Sec.~\ref{sec-halo}, the model of halo
should in principle include its finite lifetime. However, the finite 
lifetime increases the halo size in the out direction and the 
`free' correlation function in out direction is distorted at even 
smaller momenta than the `free' correlation functions in side and 
long directions. Therefore, our conclusions cannot be changed by 
taking into account a finite duration of pion emission from the halo. 


\section{Conclusions}


Let us summarize our study of the two-particle correlation functions. 
We have derived a relativistic generalization of the nonrelativistic 
Koonin formula. The calculations have been performed in the 
center-of-mass frame of the pair where a nonrelativistic wave function 
of the particle's relative motion is meaningful. It required an 
explicit transformation of the source function to the center-of-mass 
frame of the pair. Finally, the correlation function has been 
transformed to the source rest frame as a Lorentz scalar field.
The Coulomb correlation functions of pairs of identical pions and 
kaons have been computed. The source has been anisotropic and of 
finite lifetime. For pions the effect of halo has been also taken 
into account. The source function has been always of the Gaussian 
form. 
 
Having the exact Coulomb correlation functions, the Bowler-Sinyukov
procedure to remove Coulomb effect was tested. It was shown that
the procedure works very well even for an extremely anisotropic 
source provided the halo is absent. For kaons small deviations are 
observed for a sufficiently large source. When the halo is included 
the pion correlation function are noticeably distorted for very small
relative momenta but the source radii remain uninfluenced. Thus,
we conclude that the Bowler-Sinyukov procedure, which at first glance
does not look very reliable, appears to be surprisingly accurate. 
A possible interplay of Coulomb effects and fireball's expansion 
has not been studied here but our analysis shows that Coulomb effects
are not sensitive to the source's shape as long as the characteristic 
source radius is much smaller than the Bohr radius of the particle's 
pair of interest. Then, the Bowler-Sinyukov procedure is expected
to work well.  

\appendix


\section{Derivation of Correlation Function}


We sketch here the derivation, which is discussed in detail in  \cite{Lednicky:1981su,Lednicky:1996hp,Lednicky:2005tb},
of the correlation function of two identical interacting bosons.
Under rather general conditions, the correlation function
as defined by Eq.~(\ref{def-corr-fun}) can be written down as
\be
\label{start-corr}
C(p_1,p_2) = \int d^4 x_1 d^4 x_1' 
d^4 x_2 d^4 x_2' \rho(x_1,x_2;x_1',x_2') \,
\Psi_{p_1,p_2}(x_1,x_2) \, \Psi_{p_1,p_2}^*(x_1',x_2') \;,
\ee
where $\rho(x_1,x_2;x_1',x_2')$ is the properly normalized
coordinate space density matrix describing production process 
of the two particles and $\Psi_{p_1,p_2}(x_1,x_2)$ 
is the Bethe-Salpeter amplitude; $x_1$, $x_2$, $x_1'$, $x_2'$,
$p_1$, $p_2$ are all four-vectors.

To separate the relative from center-of-mass motion, one uses 
the variables (\ref{relative}) and expresses the Bethe-Salpeter 
amplitude as $\Psi_{p_1,p_2}(x_1,x_2) = 
{\rm e}^{{\rm i}PX} \psi_q(x)$. Then, the integrals over
$X$ and $X'$ are performed and the formula (\ref{start-corr}) 
changes into 
\be
\label{start-corr-relat}
C(P,q) = \int d^4 x \, d^4 x' 
\rho_P(x;x') \, 
\psi_q(x) \, \psi_q^*(x') \;.
\ee
And now one argues that the density matrix $\rho_P(x;x')$
can be approximated by the diagonal form
\be
\label{rho-diagonal}
\rho_P(x;x') = D_r(x) \: \delta^{(4)}(x-x') \;,
\ee
where $D_r(x)$ is the relative source function of probabilistic
interpretation. The nonrelativistic counterpart of $D_r(x)$
is given by Eq.~(\ref{source-relat}). To justify the expression
(\ref{rho-diagonal}) one assumes that the effect of particle 
production can be factorized from the final state interaction,
as the production process occurs at a much larger energy-momentum
scale than the process of final state interaction. Substituting
the formula (\ref{rho-diagonal}) into Eq.~(\ref{start-corr-relat})
one finds
\be
\label{corr-final}
C(P,q) = \int d^4 x \, D_r(x) \, |\psi_q(x)|^2 \;.
\ee
When the Bethe-Salpeter amplitude $\psi_q(x)$ is
transformed to the center-of-mass frame of the pair of particles,
it can be replaced by the nonrelativistic function 
$\varphi_\mathbf{q_*}(\mathbf{r_*})$ when $t_*$ is assumed
to vanish. Then, one reproduces our formula (\ref{Koonin-relat}).


\section{Approximation of Confluent Hypergeometric Function}


We derive here an approximate expression of the Coulomb scattering
function which holds when the source size is much smaller than the
Bohr radius of the two interacting particles. The confluent 
hypergeometric function $_1 F_1(a,b;z)$, which gives the
Coulomb scattering function, is defined as 
\be
\label{confluent1}
_1 F_1(a,b;z) = 1 + \sum^{\infty}_{n\;=\;1}\frac{z^n}{n!}
\prod^{n-1}_{k\,=\,0}\frac{a+k}{b+k} \;.
\ee
The Coulomb scattering function corresponds to the arguments 
$a = -i\eta/q$, $b = 1$ and $z = i(qr-\mathbf{qr})$. Introducing 
the parabolic coordinate $\xi$ we have $q\xi=qr-\mathbf{qr}$ and 
thus, 
\be
\label{confluent2}
_1 F_1\Big(-\frac{i\eta}{q},1;iq\xi \Big) = 
1+\sum^{\infty}_{n\,=\,1}\frac{(iq\xi)^n}{(n!)^2}
\prod^{n-1}_{k\,=\,0}(-\frac{i\eta}{q}+k) \;.
\ee

To obtain the desired approximation we write down a few first terms 
of the series (\ref{confluent2}) and we rearrange them as
\ba
\nonumber 
_1 F_1 \Big(-\frac{i\eta}{q},1;iq\xi \Big) 
&=& 
1+\frac{-\frac{i\eta}{q}iq\xi}{(1!)^2}+
\frac{-\frac{i\eta}{q}(-\frac{i\eta}{q}+1)(iq\xi)^2}{(2!)^2}
+\frac{-\frac{i\eta}{q}(-\frac{i\eta}{q}+1)
(-\frac{i\eta}{q}+2)(iq\xi)^3}{(3!)^2}+ \dots
\\[2mm] \nonumber
&=&
1+\frac{\eta \xi}{(1!)^2}
+\frac{\eta \xi (\eta \xi + iq\xi )}{(2!)^2}
+\frac{\eta \xi (\eta \xi + iq\xi )(\eta \xi + 2iq\xi )}{(3!)^2}
+ \dots
\\ [2mm]
\label{baza2}
&=&
1 + \sum^{\infty}_{n\,=\,1}\frac{1}{(n!)^2}
\prod^{n-1}_{k\,=\,0}(\eta \xi+kiq\xi).
\ea

\begin{figure}[t]
\begin{minipage}{8cm}
\centering
\includegraphics*[width=8cm]{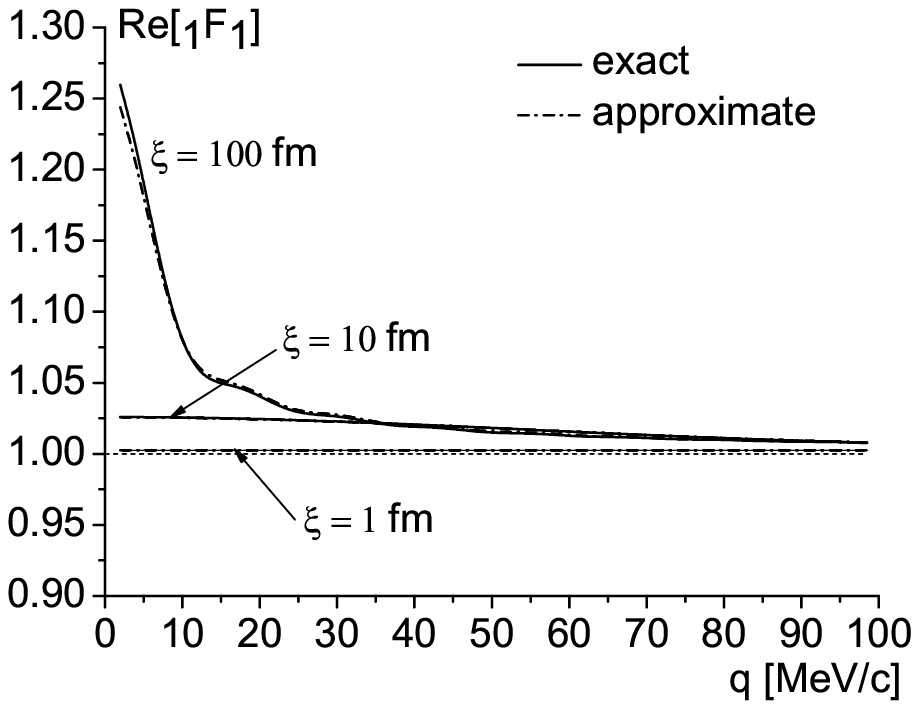}
\end{minipage}\hspace{3mm}
\centering
\begin{minipage}{8cm}
\includegraphics*[width=8cm]{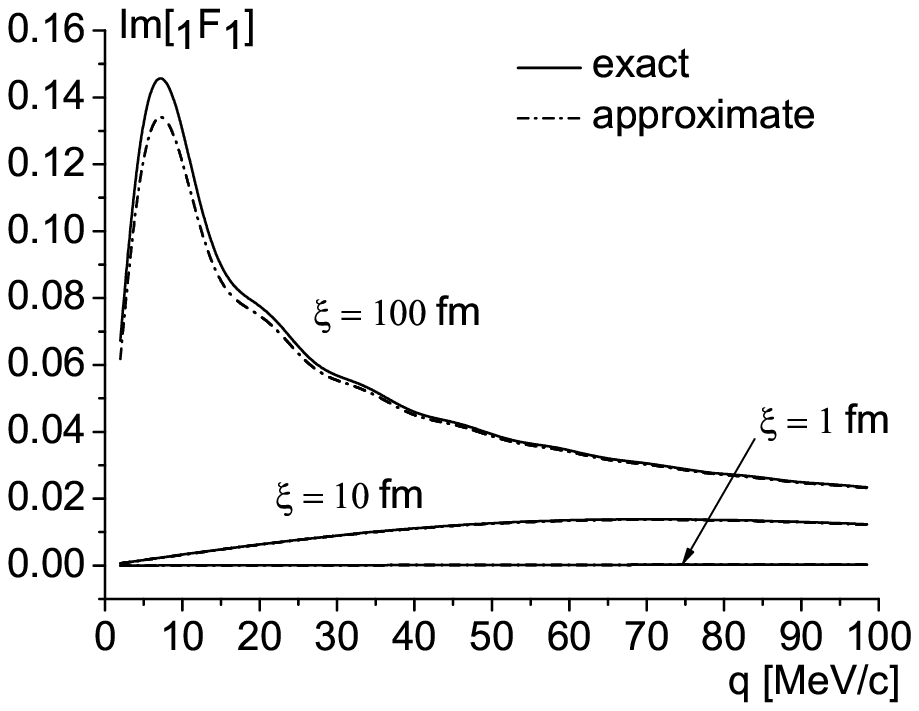}
\end{minipage}
\caption{The real (left panel) and imaginary (right panel) part
of the pion Coulomb function $_{1}F_1(-i\eta_\pi/q,1;iq\xi)$
as a function of $q$ for several values of $\xi$. The solid and dashed
lines represent, respectively, the exact formula (\ref{confluent2}) 
and the approximation (\ref{approx-1}).}
\label{fig-1F1}
\end{figure}

We first note that with the formula (\ref{baza2}), one easily finds 
the value of the Coulomb function at $q = 0$
\be
\label{baza3}
_1F_1\Big(-\frac{i\eta}{q},1;iq\xi \Big)\Big|_{q=0}
= 1 + 
\sum^{\infty}_{n\,=\,1}\frac{1}{(n!)^2}\prod^{n-1}_{k\;=\;0}(\eta \xi)
= 
\sum^{\infty}_{n\,=\,0}\frac{(\eta \xi)^n}{(n!)^2}=I_0(2\sqrt{\eta \xi}) \:,
\ee
where $I_v(z)$ is the modified Bessel function of the first 
kind defined as 
$$
I_v(z)=\sum^{\infty}_{k\,=\,0}
\frac{1}{\Gamma(k+v+1)k!}\left(\frac{z}{2}\right)^{2k+v} \;.
$$

We define the new variables $x \equiv \eta \xi$ and $y \equiv iq/\eta$,
and we write down the series (\ref{baza2}) as
\be
\label{baza5}
_1F_1\Big(\frac{1}{y},1;xy \Big) = 1 
+ \underbrace{\frac{x}{(1!)^2}}_{n=1}
+ \underbrace{\frac{x^2(1+y)}{(2!)^2}}_{n=2}
+ \underbrace{\frac{x^3(1+y)(1+2y)}{(3!)^2}}_{n=3}+ \dots 
\ee
We are interested in the approximation which holds when the source 
size is much smaller than the Bohr radius of the scattering
particles that is when $x \ll 1$. Since $y$ can be arbitrary big,
the series cannot be simply terminated at a given power of $x$. 
Instead, one should take into account the lowest power of $x$ for 
every power of $y$. For this purpose we have to rearrange
the series (\ref{baza5}). After rather tedious analysis, one shows 
that 
\be
\label{baza6}
_1F_1\Big(\frac{1}{y},1;xy \Big) = 
1+\sum^{\infty}_{k\,=\,0} y^k 
\sum^{\infty}_{n\;=\;k+1}
\left(\sum^{n-k}_{l_1\,=\,1}
\sum^{n-k+1}_{l_2\,=\,l_1+1} \dots 
\sum^{n-1}_{l_k\,=\,l_1+k-1}l_1l_2 \dots l_k\right)
\frac{x^n}{(n!)^2} \;.
\ee
And now for each $k$ in the series (\ref{baza6}) we take into account 
only the term of the lowest order of $x$ that is we include only the
term of $n=k+1$. Observing that
$$
\sum^{1}_{l_1\,=\,1}
\sum^{2}_{l_2\,=\,l_1+1} \dots 
\sum^{k}_{l_k\,=\,l_1+k-1}l_1l_2 \dots l_k = k! \;,
$$
we obtain the desired approximation
\be
\label{baza7}
_1F_1\Big(\frac{1}{y},1;xy \Big) \approx
1 +   
\frac{1}{y}\sum^{\infty}_{k\,=\,0} \frac{(xy)^{k+1}}{(k+1)!(k+1)}
=
1 + \frac{i}{y} {\rm Si}(-ixy) - \frac{1}{y}
\big( \gamma_e + \ln (-ixy)- {\rm Ci}(-ixy)\big) \;,
\ee
where ${\rm Si}(z)$ and ${\rm Ci}(z)$ are integral sine and cosine
functions, respectively, and $\gamma_e \approx 0.5772$ is the Euler's constant.
Reintroducing the physical arguments, we finally have
\be
\label{approx-1}
_1 F_1\Big(-\frac{i\eta}{q},1;iq\xi \Big) 
\approx
1 + \frac{\eta}{q} {\rm Si}(q\xi) + i\frac{\eta}{q}
\big( \gamma_e + \ln (q\xi)- {\rm Ci}( q\xi)\big) \;.
\ee

\begin{figure}[t]
\begin{minipage}{8cm}
\centering
\includegraphics*[width=8cm]{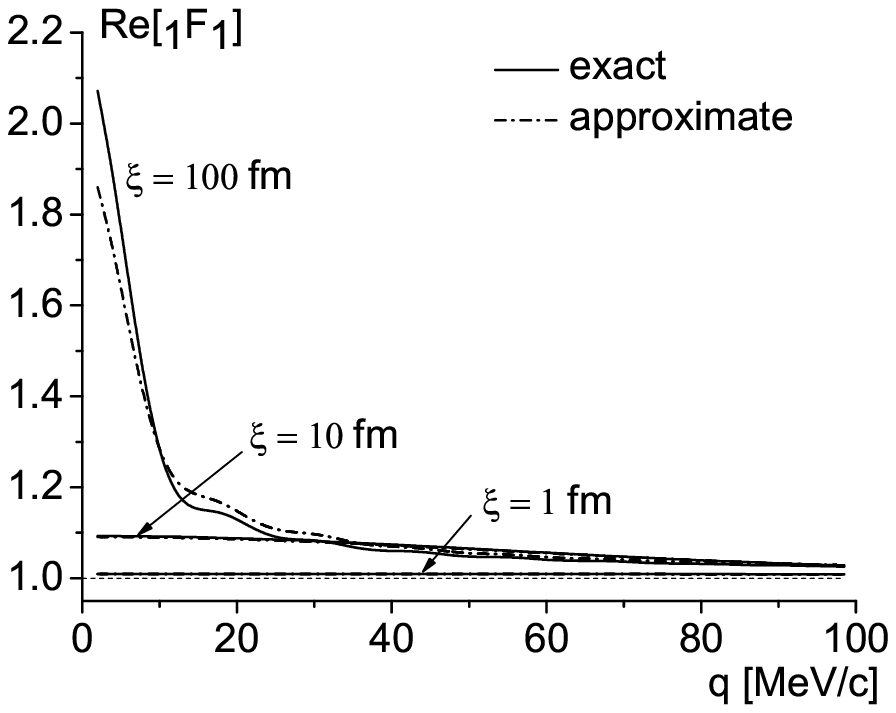}
\end{minipage}\hspace{3mm}
\begin{minipage}{8cm}
\centering
\includegraphics*[width=8cm]{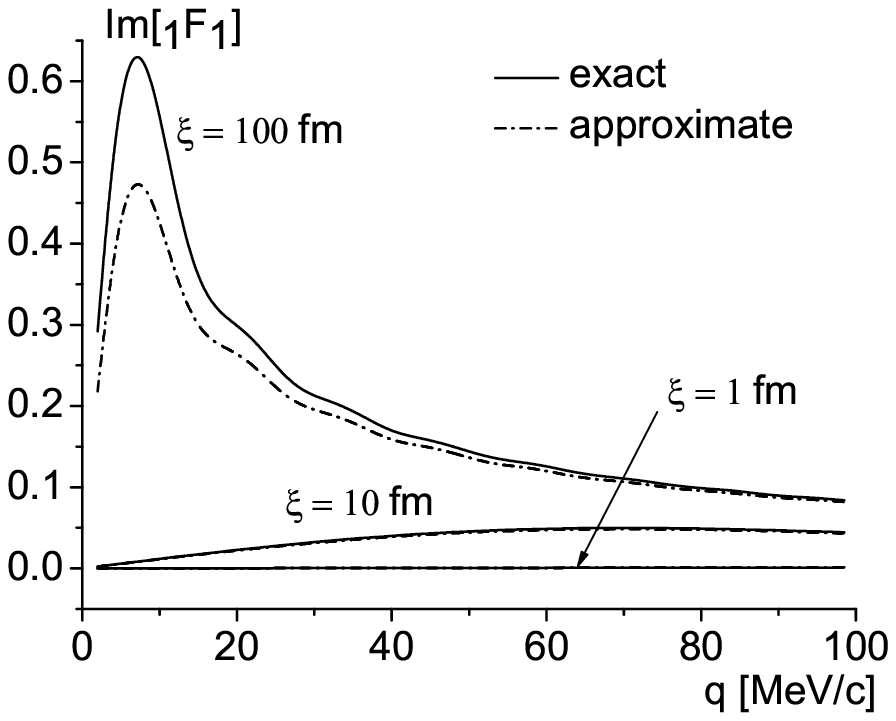}
\end{minipage}
\caption{The real (left panel) and imaginary (right panel) part
of the kaon Coulomb function $_{1}F_1(-i\eta_K/q,1;iq\xi)$
as a function of $q$ for several values of $\xi$. The solid and dashed
lines represent, respectively, the exact formula (\ref{confluent2}) 
and the approximation (\ref{approx-1}).}
\label{fig-1F1-KK}
\end{figure}

In Figs.~\ref{fig-1F1} and \ref{fig-1F1-KK} we show the Coulomb functions
computed from the approximate formula (\ref{approx-1}) for the Bohr radius
of pions and kaons, respectively. As seen, the approximation works very well 
for pions ($\eta_\pi^{-1} = 388 \;{\rm fm}$) but it is not so accurate for 
kaons ($\eta_K^{-1} = 110 \;{\rm fm}$). We also see that 
${\rm Re}[ _{1}F_1(-i\eta /q,1;iq\xi)] 
\gg {\rm Im} [ _{1}F_1(-i\eta /q,1;iq\xi)]$.
Therefore, the imaginary part can be neglected and 
\be
\label{approx-2}
_1F_1\Big(-\frac{i\eta}{q},1;iq\xi \Big) \approx
1 + \frac{\eta}{q} {\rm Si}(q\xi) \;.
\ee
Since $\frac{\eta}{q} {\rm Si}(q\xi) \ll 1$, we also have
\be
\label{approx-3}
\Big|{_1}F_1\Big(-\frac{i\eta}{q},1;iq\xi \Big)\Big|^2 
\approx
1 + 2\frac{\eta}{q} {\rm Si}(q\xi) \;.
\ee
The approximations (\ref{approx-2}, \ref{approx-3}) were used to 
compute the correlation functions of pions but not of kaons.

\section*{Acknowledgements}

We are very grateful to W.~Broniowski, W.~Florkowski, A.~Kisiel 
and R.~Lednick\'y for numerous fruitful discussions. This work was 
partially supported by Polish Ministry of Science and Higher Education 
under grant N202 080 32/1843.


\end{document}